\documentclass[nofootinbib,aps,pra,superscriptaddress,reprint]{revtex4-2}
\usepackage{amsmath,amsthm,amsfonts,amssymb}
\usepackage{newtx}
\usepackage{graphicx}
\usepackage{bm}
\usepackage{braket} 
\usepackage{fancyhdr}
\usepackage{soul} 
\usepackage{times} 
\usepackage{color} 
\usepackage[colorlinks,linkcolor=blue,anchorcolor=blue,citecolor=blue,urlcolor=black]{hyperref}
\newcommand{\bigchi}{\makebox{\large\ensuremath{\chi}}}
\DeclareMathAlphabet{\mathcal}{OMS}{cmsy}{m}{n}

\makeatletter
\newcommand{\extp}{\@ifnextchar^\@extp{\@extp^{\,}}}
\def\@extp^#1{\mathop{\bigwedge\nolimits^{\!#1}}}
\makeatother

\begin{document}

\title{Performance optimization of continuous variable quantum teleportation with\\ generalized photon-varying non-Gaussian operations}

\date{\today}

\author{Mingjian He}
\email{mingjian.he@ccnu.edu.cn}

\author{Shouyin Liu}
\email{syliu@ccnu.edu.cn}

\affiliation{College of Physical Science and Technology,\\
			Central China Normal University, Wuhan, Hubei 430079, China.}

\begin{abstract}
Continuous variable quantum teleportation provides a path to the long-distance transmission of quantum states.
Photon-varying non-Gaussian operations have been shown to improve the fidelity of quantum teleportation when integrated into the protocol.
However, given a fixed non-Gaussian operation, the achievable fidelity varies with different input states.
An operation that increases the fidelity for teleporting one class of states might do the contrary for other classes of states.
A performance metric suitable for different input states is missing.
For a given type of non-Gaussian operation, the achievable fidelity also varies with parameters associated with the operation. 
Previous work only focuses on particular settings of the parameters.
Optimization over the parameters is also missing.
In this work, we build a framework for photon-varying non-Gaussian operations for multi-mode states, upon which we propose a performance metric suitable for arbitrary teleportation input states.
We then apply the new metric to evaluate different types of non-Gaussian operations. 
Starting from simple multi-photon photon subtraction and photon addition, we find that increasing the number of ancillary photons involved in the operation does not guarantee performance improvement.
We then investigate combinations of the operations mentioned above, finding that operations that approximate a particular form provide the best improvement.
The results provided here will be valuable for real-world implementations of quantum teleportation networks and applications that harness the non-Gaussianity of quantum states.

\end{abstract}

\maketitle

\thispagestyle{fancy}
\pagestyle{fancy}
\renewcommand{\headrulewidth}{0pt}

\section{Introduction}

Continuous-variable (CV) quantum teleportation protocols, adaptable with off-the-shelf optical technologies, provide a path to the long-distance transmission of quantum information encoded in quantum states.
Many breakthroughs have been made in the deployment of large-scale quantum teleportation systems over the past decades, e.g., \cite{yonezawa2004demonstration,jin2010experimental,ma2012quantum,ren2017ground,lago2023long,zhao2023enhancing}.

The core of teleportation is the consumption of entangled \textit{resource states} distributed before the protocol begins.
However, imperfect resource states caused by finite power during state production and channel loss during state distribution limit the achievable fidelity of the \textit{input state} (the state to be teleported).
One class of resource states is the two-mode squeezed vacuum (TMSV) state, which can be produced by combining two single-mode squeezed vacuum states at a balanced beam splitter.
Perfect teleportation requires TMSV states with infinite squeezing, which is unachievable in theory and experiment (e.g., 15 dB squeezing for state-of-the-art technologies \cite{vahlbruch2016detection}).
TMSV states with higher squeezing are more sensitive to decoherence during distribution over lossy channels \cite{mivcuda2012noiseless}, making the long-distance distribution of TMSV states another outstanding challenge.

Non-Gaussian operations have been shown to enhance entanglement and other desirable properties when applied to Gaussian states.
Much attention have been drawn to the studies of such operations in various quantum information tasks, such as quantum key distribution \cite{huang2013performance,borelli2016quantum,li2016non-gaussian,guo2017performance,zhao2017improvement,ma2018continuous-variable,guo2019continuous-variable,PhysRevA.102.012608,he2019multi,singh2021non,chen2023continuous,wang2023non},
noiseless linear amplification \cite{ralph2009nondeterministic,kim2012quantum,yang2012continuous-variable,gagatsos2014heralded,zhang2018photon,adnane2019quantum,2022_443}, 
entanglement distillation \cite{lund2009continuous-variable,zhang2010distillation,fiurasek2010distillation,datta2012compact,lee2013entanglement,zhang2013continuous-variable-entanglement,seshadreesan2019continuous-variable,mardani2020continuous}, 
and quantum computing \cite{ghose2007non,marshall2015repeat,miyata2016implementation,marshall2016continuous}.
In teleportation, non-Gaussian operations can improve fidelity at the cost of reduced overall efficiency, e.g.,  \cite{2002_188,2007_189,2009_183,2010_333,2015_297,2015_325,2015_334,2017_178,ye2020nonclassicality,2020_485,2021_380,villasenor2021enhancing,2022_483,2023_475,2023_478,2023_486,2023_489}.
However, previous work focuses on the fidelity of particular classes of input states, e.g., coherent states, squeezed states, and CV qubit states.
The achievable fidelity varies with different input states for a given resource state. 
For example, for coherent states input, applying photon subtraction to the resource states increases fidelity \cite{2002_188} while using photon addition decreases the fidelity \cite{2007_189}.
On the contrary, for squeezed states input, applying photon subtraction and photon addition both increase the fidelity \cite{2009_183}.
The energy-constrained channel fidelity was proposed in \cite{2022_496} and developed in \cite{2023_494} as a performance metric for teleportation systems.
However, the metric is independent of the input states and does not indicate the achievable fidelity for any input states.
A performance metric suitable for different input states is still missing.

For a given type of non-Gaussian operation, the achievable fidelity also varies with parameters associated with the operation.
For example, varying the number of photons being subtracted (or added) in photon subtraction (or addition) provides different fidelity \cite{2015_297,2015_334,2015_325,2017_178}.
The fidelity offered by a cascaded application of the same operation also differs as the number of operations varies \cite{2023_486}.
However, previous work focuses on particular parameter settings, e.g., restricting the number of ancillary photons in the operation or the number of operations to a few (mostly $\leq3$).
It remains unclear whether increasing the numbers always improves fidelity.
An optimization over the parameters of non-Gaussian operations is also missing.

This paper aims to fully reveal the potential of the photon varying (PV) operations (i.e., operations that include both photon subtraction and photon addition) in CV teleportation.
To this end, we first build a framework for PV operations in the multi-mode setting, expressing the characteristic function of the photon-varied state as a multi-variable multi-index Hermite-Gaussian function.
Based on the framework built, we introduce a performance metric suitable for arbitrary input states.
We then propose generalized PV operations with parameters optimized using the new metric as the cost function.
Our results show that the optimal generalized PV operations approach a specific Gaussian operation, the noiseless linear amplification.

The remaining sections are organized as follows. Section~\ref{sec:2} establishes a framework for CV teleportation with TMSV resource states and photon varying operations, upon which the generalized photon varying operations are introduced and investigated in Section~\ref{sec:3}. Section~\ref{sec:4} studies the above operations in teleportation with other resource states.
Conclusions of the paper are given in Section~\ref{sec:5}.

\section{CV teleportation with photon varying operations}\label{sec:2}
We consider the VBK (Vaidman-Braunstein-Kimble) protocol, which is the first CV teleportation protocol proposed in \cite{1994_366} and developed in \cite{1998_302}.
In the VBK protocol, Alice first prepares a TMSV state with modes 1 and 2 and sends mode 2 to Bob.
The distributed state is then used as the resource state for teleportation.
Alice couples the input mode (to which the input state is encoded) with mode 1 at a 50:50 beam splitter.
The $p$-quadrature of one output of the beam splitter and the $q$-quadrature of the other output are measured by two homodyne detectors.
The measurement outcome is sent to Bob through a lossless classical channel.
Based on the outcome, Bob applies a displacement operation to mode 2.
In the limit of infinite squeezing of the resource state, mode 2 approaches the input mode.

We define the Wigner characteristic function (CF) of a single-mode state with density operator $\hat{\rho}$ as
\begin{equation}
	\bigchi(\xi)=\mathrm{tr}\{\hat{\rho}\hat{D}(\xi)\},
\end{equation}
where $\xi\in\mathbb{C}$,
$\mathrm{tr}\{\cdot\}$ represents the trace operation,
$\hat{D}(\xi)=e^{\xi\hat{a}^\dagger-\xi^*\hat{a}}$ is the displacement operator, $\hat{a}$ and $\hat{a}^\dagger$ are the ladder operators of $\hat\rho$, and $(\cdot)^*$ represents complex conjugate.
The relation between the input and output modes of the VBK protocol can be described using the CF formalism \cite{marian2006continuous}
\begin{equation}\label{eq:tpinandout}
	\bigchi_{\mathrm{out}}(\xi) =
	\bigchi(\xi, \xi^*)
	\bigchi_{\mathrm{in}}(\xi),
\end{equation}
where $\bigchi_{\mathrm{in}}(\xi)$ is the CF of the input mode, $\bigchi_{\mathrm{out}}(\xi)$ is the CF of the output mode (averaged over Alice's measurement outcomes), and $\bigchi(\xi_1, \xi_2)$ is the CF of the resource state.
The above equation indicates that the protocol can be viewed as a quantum channel with a $\xi$-dependent \textit{response function} $\bigchi(\xi, \xi^*)$.
The output mode after passing the quantum channel approaches the input mode when the response function is flat, i.e., $\bigchi(\xi, \xi^*)\rightarrow1$ for all $\xi$.
The performance of teleportation can be measured by the fidelity between the input mode and the output mode, which quantifies the closeness of the CFs of the two modes \cite{chizhov2002continuous}
\begin{equation}\label{eq:fidelitycf}
	\mathcal{F} = \frac{1}{\pi} \int d^2 \xi 
	\bigchi_\mathrm{in}(-\xi) 
	\bigchi_\mathrm{out}(\xi),
\end{equation}
where $d^2\xi :=d\xi\extp d\xi^*$ with $\extp$ the exterior product.
The fidelity approaches unit for any input states when the response function $\bigchi(\xi, \xi^*)\rightarrow1$.

\begin{figure}
	\centering
	\includegraphics[width=1\linewidth]{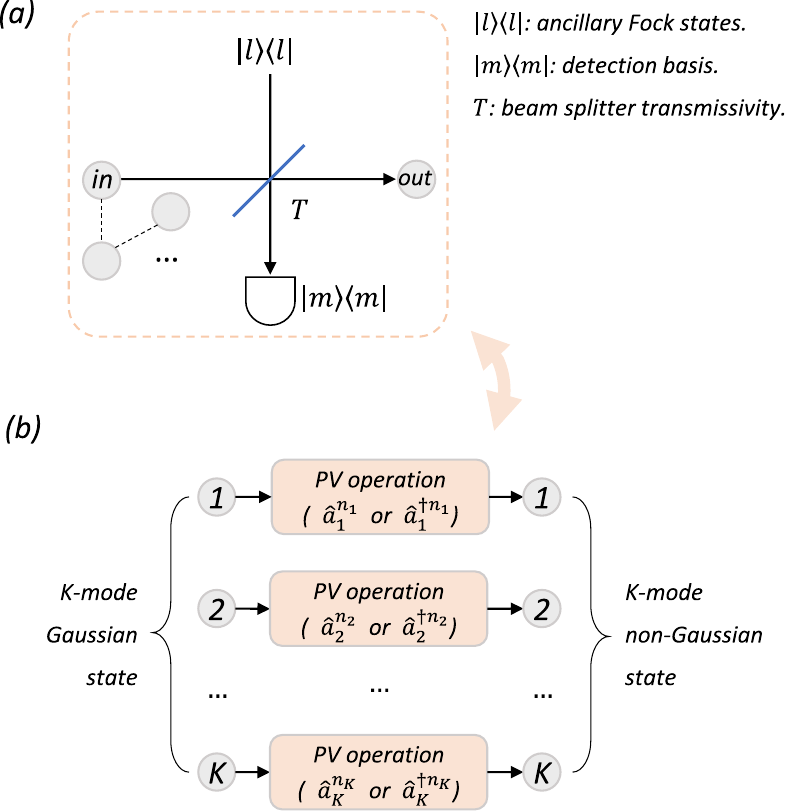}
	\caption{(a) A physical implementation to the PV operations. 
		(b) The application of $K$ PV operations to a $K$-mode Gaussian state.}
	\label{fig:diag1}
\end{figure}

Photon subtraction ($\hat{a}^n,\ n\in\mathbb{N}$) and photon addition ($\hat{a}^{\dagger n},\ n\in\mathbb{N}$) are two types of non-Gaussian operations, both shown to improve the teleportation fidelity for certain input states.
In \cite{guerrini2023photon} the states applied by the two operations mentioned above are named as photon-varied quantum states.
In this paper, we will use photon-varying (PV) operations to refer to the two operations.
A physical implementation to the PV operations is shown in Fig.~\ref{fig:diag1}a, which consists of an ancillary Fock state $\ket{l}$, a photon number resolving detector, and a beam splitter with transmissivity $T<1$.
In the operation, the initial mode is coupled with the ancillary state at the beam splitter. A photon number resolving detection is performed on one output of the beam splitter. The operation has been successful if $m$ photons are detected.
Photon subtraction $\hat{a}^n$ can be approximated by setting $l=0$, $m=n$, and $T\rightarrow1$ while photon addition $\hat{a}^{\dagger n}$ can be approximated by setting $l=n$, $m=0$, and $T\rightarrow1$.
Both operations can also be implemented by a cascaded applications of the same single-photon operations.

The the transformation for performing a PV operation on the single-mode state $\hat{\rho}$ can be written as
\begin{equation}\label{eq:nGrho}
	 \hat{\rho}\rightarrow\hat\rho_{\mathrm{PV}}=
	 \frac{1}{\mathrm{tr}\{\hat{a}^n\hat{\rho}\hat{a}^{\dagger n}\}}
	 \hat{a}^n\hat{\rho}\hat{a}^{\dagger n},
\end{equation}
where $\hat{a}$ is replaced with $\hat{a}^\dagger$ for photon addition.
Using the identities
\begin{equation}
	\begin{aligned}			
	&\mathrm{tr}\{\hat{a}\hat{\rho}\hat{D}(\xi)\}=
	-e^{-\frac{|\xi|^2}{2}}\frac{\partial}{\partial \xi^*}
	\left[e^{\frac{|\xi|^2}{2}}\bigchi(\xi)\right],\\
	&\mathrm{tr}\{\hat{\rho}\hat{a}^\dagger\hat{D}(\xi)\}=
	e^{-\frac{|\xi|^2}{2}}\frac{\partial}{\partial \xi}
	\left[e^{\frac{|\xi|^2}{2}}\bigchi(\xi)\right],\\
	&\mathrm{tr}\{\hat{a}^\dagger\hat{\rho}\hat{D}(\xi)\}=
	e^{\frac{|\xi|^2}{2}}\frac{\partial}{\partial \xi}
	\left[e^{-\frac{|\xi|^2}{2}}\bigchi(\xi)\right],\\
	&\mathrm{tr}\{\hat{\rho}\hat{a}\hat{D}(\xi)\}=
	-e^{\frac{|\xi|^2}{2}}\frac{\partial}{\partial \xi^*}
	\left[e^{-\frac{|\xi|^2}{2}}\bigchi(\xi)\right],
\end{aligned}
\end{equation}
the transformation in Eq.~(\ref{eq:nGrho}) can be written in an equivalent CF form as
\begin{equation}\label{eq:partialchi}
	\bigchi(\xi)\rightarrow
	\bigchi_{\mathrm{PV}}(\xi)=\frac{1}{\mathcal{N}}O_{t,n}(\xi)\bigchi(\xi),
\end{equation}
where
\begin{equation}
	O_{t,n}(\xi)=(-1)^ne^{t\frac{|\xi|^2}{2}}
	\frac{\partial^{2n}}{\partial\xi^n\partial\xi^{*n}}
	e^{-t\frac{|\xi|^2}{2}},
\end{equation}
$\mathcal{N}=O_{t,n}(\xi)\bigchi(\xi)\big|_{\xi=0}$ is the normalization constant, and $t=-1$ for photon subtraction and $t=1$ for photon addition.
In \cite{guerrini2023photon} it is shown that the operator $O_{t,n}(\xi)$ can be simplified to a generalized two-variable Hermite polynomial function when the initial state is a \textit{single-mode} Gaussian state.
To find the simplified form of the CF of the TMSV state after the PV operation, we first extend the result in \cite{guerrini2023photon} to the general multi-mode case.

For a symmetric $2K$-by-$2K$ matrix $\bm{M}$, we define the generalized multi-index multi-variable Hermite function by the generating function
\begin{equation}\label{eq:hermitedef}
	\sum_{n_1,...,n_{2K}=0}^\infty
	\left(\prod_{i=1}^{2K}\frac{u_i^{n_i}}{n_i!}\right)
	H_{n_1,...,n_{2K}}(\bm{x};\bm{M})=
	e^{{\bm{u}^T\bm{M}\bm{u}}+\bm{x}^T\bm{u}},
\end{equation}
where $\bm{x},\bm{u}\in\mathbb{C}^{2K}$ are column vectors, the $u_i$'s are elements of $\bm{u}$, and $(\cdot)^T$ represents matrix transpose (not conjugate transpose).
First combining the terms of the Taylor expansion for the right side of Eq.~(\ref{eq:hermitedef}), performing a change of summation indices, and then comparing the terms, leads to the analytical solution to $H_{n_1,...,n_{2K}}(\bm{x};\bm{M})$,
\begin{equation}\label{eq:hexpression}
	\begin{aligned}
	&H_{n_1,...,n_{2K}}(\bm{x};\bm{M})\\
	&\hskip 10pt=\sum_{\{n_{i,j}\}}	\left[\prod_{i=1}^{2K}	
	\frac{{n_i!}{x_i}^{q_i}}{q_i!}
	\frac{M_{i,i}^{n_{i,i}}}{n_{i,i}!}
	\prod_{j=i+1}^{2K}
	\frac{(2M_{i,j})^{n_{i,j}}}{n_{i,j}!}\right],
	\end{aligned}
\end{equation}
where the $M_{i,j}$'s are elements of the matrix $\bm{M}$, 
the $x_{i}$'s are elements of the vector $\bm{x}$, 
the summation is taken over all possible combinations of the $n_{i,j}$'s ($1\leq i\leq 2K,\ i\leq j\leq 2K$) such that $q_i\geq0$ for all $i$,
and
\begin{equation}
q_i=n_i-n_{i,i}-\sum_{i'=1}^{2K} n_{\min(i,i'), \max(i,i')}.
\end{equation}

From Eq.~(\ref{eq:hermitedef}) it follows that for a column vector $\bm{d}\in\mathbb{C}^{2K}$,
\begin{equation}
	\begin{aligned}		
	&\sum_{n_1,\dots,n_{2K}=0}^\infty\left(\prod_{i=1}^{2K}\frac{u_i^{n_i}}{n_i!}\right)
	H_{n_1,\dots,n_{2K}}(\bm{M}\bm{x}-\bm{d};-\frac{1}{2}\bm{M})\\
	&\hskip 20pt\times
	e^{-\frac{1}{2}\bm{x}^T\bm{M}\bm{x}+\bm{d}^T\bm{x}}
	\\
	&\hskip 10pt=
	e^{-\frac{1}{2}(\bm{x}-\bm{u})^T\bm{M}(\bm{x}-\bm{u})+\bm{d}^T(\bm{x}-\bm{u})}.
\end{aligned}
\end{equation}
Comparing the terms in the Taylor expansion for the right side of the above equation it follows that
\begin{equation}\label{eq:partialh}
	\begin{aligned}	
	&\frac{\partial^{\sum_{i=1}^{2K}n_i}}{\prod_{i=1}^{2K}(\partial x_i)^{n_i}}
	e^{-\frac{1}{2}\bm{x}^T\bm{M}\bm{x}+\bm{d}^T\bm{x}}\\
	&\hskip10pt=(-1)^{\sum_{i=1}^{2K}n_i}
	H_{n_1,\dots,n_{2K}}(\bm{M}\bm{x}-\bm{d};-\frac{1}{2}\bm{M})
	e^{-\frac{1}{2}\bm{x}^T\bm{M}\bm{x}+\bm{d}^T\bm{x}}.
\end{aligned}
\end{equation}

A $K$-mode Gaussian state with covariance matrix $\bm{V}$ and vector of means $\bm{\mu}$ ($\bm{V}$ and $\bm{\mu}$ are both for the quadratures) has the CF
\begin{equation}
	\bigchi(\bm\xi)=e^{-\frac{1}{2}\tilde{\bm\xi}^\dagger
	\tilde{\bm{V}}\tilde{\bm\xi}
	+\tilde{\bm{\mu}}^\dagger\tilde{\bm\xi}},
\end{equation}
where 
$\bm\xi=[\xi_1,\dots,\xi_K]\in\mathbb{C}^{K}$, 
$\tilde{\bm\xi}=[\xi_1, {\xi_1}^*,\dots,\xi_K,{\xi_K}^*]^\dagger$ is the augmented vector of $\bm\xi$, 
$\tilde{\bm{V}}=\bm{Z}\bm{J}\bm{V}\bm{J}^\dagger\bm{Z}$, 
$\tilde{\bm{\mu}}=\bm{Z}\bm{J}\bm{\mu}$, 
$\bm{Z}=\bm{I}_K\otimes[1,0;0,-1]$, 
$\bm{J}=\bm{I}_K\otimes[1,i;1,-i]/2$, 
$\bm{I}_K$ is the $K$-by-$K$ identity matrix,
and we have set the variance of the vacuum state to 1 shot noise unit (i.e., $\hbar=2$) and will use this setting throughout the paper.
As shown in Fig.~\ref{fig:diag1}b, suppose every mode of the $K$-mode state undergoes an independent PV operation with $n_i$ ($1\leq i\leq K$) photons varied.
Combining Eqs.~(\ref{eq:partialchi}) and (\ref{eq:partialh}), the CF for the state after the operations can then be written as
\begin{equation}\label{eq:pvcf}
	\begin{aligned}			
	&\bigchi_{\mathrm{PV}}(\bm\xi)\\
	&\hskip 10pt=\frac{(-1)^{\sum_{i=1}^{K}n_i}}{\mathcal{N}}
	H_{n_1,n_1,\dots,n_{K},n_{K}}
	\left(\bm{X}(\tilde{\bm{V}}'
	\tilde{\bm\xi}+\tilde{\bm{\mu}});-\frac{1}{2}\bm{X}\tilde{\bm{V}}'\right)\\
	&\hskip 20pt\times\bigchi(\bm\xi),
\end{aligned}
\end{equation}
where $\tilde{\bm{V}}'=\tilde{\bm{V}}+\mathrm{diag}[t_1,t_1,\dots,t_K,t_K]/2$, 
$t_i=-1$ (or $1$) when photon subtraction (or photon addition) is applied to the $i$-th mode,
$\bm{X}=\bm{I}_K\otimes[0,1;1,0]$, 
and
\begin{equation}
\mathcal{N}=(-1)^{\sum_{i=1}^{K}n_i}H_{n_1,n_1,\dots,n_{K},n_{K}}
\left(\bm{X}\tilde{\bm{\mu}};-\frac{1}{2}\bm{X}\tilde{\bm{V}}'\right).	
\end{equation}
For conciseness,
we will use 
$H_{n_1,n_1,\dots,n_{K},n_{K}}(\bm\xi)$
to represent
$H_{n_1,n_1,\dots,n_{K},n_{K}}
\left(\bm{X}(\tilde{\bm{V}}'
\tilde{\bm\xi}+\tilde{\bm{\mu}});-\frac{1}{2}\bm{X}\tilde{\bm{V}}'\right)$
in the rest of the paper when the context is clear.

We can now proceed with the two-mode case.
Applying PV operations (with parameters $n_1$ and $n_2$ for the two modes, respectively) to both modes of a two-mode Gaussian state produces a photon varied state with CF of the form
\begin{equation}
	\bigchi_{\mathrm{PV}}(\xi_1,\xi_2) =
	\frac{(-1)^{n_1+n_2}}{\mathcal{N}}H_{n_1,n_1,n_2,n_2}(\xi_1,\xi_2)
	\bigchi(\xi_1,\xi_2).
\end{equation}
For an arbitrary input state, teleportation using the above state as resource state and a resource state without any operation are equivalent to quantum channels with different response functions. 
The ratio between the two response functions can be written as
\begin{equation}
	\begin{aligned}
	\frac{\bigchi_{\mathrm{PV}}(\xi, \xi^*)}{\bigchi(\xi, \xi^*)}
	&=\frac{1}{H_{n_1,n_1,n_2,n_2}(0,0)}H_{n_1,n_1,n_2,n_2}(\xi, \xi^*)\\
	&:=\mathcal{H}(\xi).
	\end{aligned}
\end{equation}
The function $\mathcal{H}(\xi)$, which we name as the \textit{response ratio} function, can be used to indicate the impact of the PV operations. 
Recall from Eq.~(\ref{eq:tpinandout}) the output state of teleportation for the PV operation case can be written as
\begin{equation}
	\bigchi_{\mathrm{out}}(\xi) =
	\mathcal{H}(\xi)\bigchi(\xi, \xi^*)
	\bigchi_{\mathrm{in}}(\xi),
\end{equation}
for Gaussian resource state with response function $\bigchi(\xi, \xi^*)\in(0,\ 1)$,
at any point $\xi_\mathrm{p}$, $\mathcal{H}(\xi_\mathrm{p})>1$ indicates that the PV operations reduce distortion caused by the quantum teleportation channel to any input state at $\xi_\mathrm{p}$.
Applying PV operations with $\mathcal{H}(\xi)>1$ for all $\xi$ to the resource state improves the teleportation fidelity for any input state.
For PV operations with $\mathcal{H}(\xi)>1$ only for a certain area of $\xi$, the increase in fidelity depends on the input state.

We now consider the TMSV state with the form
\begin{equation}
	\begin{aligned}			
		\ket{\mathrm{TMSV}}&=\hat{S}(r)\ket{0,0}_{1,2}\\&
		=\sqrt{1-\lambda^2}\sum_{n=0}^{\infty}\lambda^n\ket{n,n}_{1,2},
	\end{aligned}
\end{equation}
where $\hat{S}(r)=e^{r(\hat{a}_1^\dagger\hat{a}_2^\dagger-\hat{a}_1\hat{a}_2)}$ is the two-mode squeezing operator and $r=\mathrm{atanh}(\lambda)$ is the squeezing parameter. The CF of the TMSV state is then given by
\begin{equation}\label{eq:cfTMSV}
	\bigchi_{\mathrm{TMSV}}(\xi_1,\xi_2)=
	e^{-\frac{1}{2}\left[V\left(|\xi_1|^2+|\xi_2|^2\right)-\sqrt{V^2-1}
		\left(\xi_1\xi_2+{\xi_1}^*{\xi_2}^*\right)\right]},
\end{equation}
where $V=\cosh(2r)$ is the variance of the distribution of the quadratures.
The response ratio function for the TMSV state with PV operations can be written as
\begin{equation}
	\mathcal{H}(\xi)=H_{n_1,n_1,n_2,n_2}(\xi, \xi^*)/H_{n_1,n_1,n_2,n_2}(0,0),
\end{equation}
where
\begin{equation}\label{eq:hermite2v}
	\begin{aligned}
		&H_{n_1,n_2,n_3,n_4}(\xi, \xi^*)\\
		&\hskip 10pt=\sum_{n_5,n_6,n_7,n_8}
		\xi^{(n_2+n_3)}
		{\xi^*}^{(n_1+n_4)}
		|\xi|^{2(-n_5-n_6-n_7-n_8)}\\
		&\hskip 20pt\times\frac{n_1!n_2!(A+C)^{n_1+n_2-2n_5-n_7-n_8}A^{n_5}}{n_5!(n_1-n_5-n_7)!(n_2-n_5-n_8)!}\\
		&\hskip 20pt\times\frac{n_3!n_4!(B+C)^{n_3+n_4-2n_6-n_7-n_8}B^{n_6}}{n_6!(n_3-n_6-n_7)!(n_4-n_6-n_8)!}\frac{C^{n_7+n_8}}{n_7!n_8!},
	\end{aligned}
\end{equation}
the summation is taken over all possible $n_i$'s ($i=5,6,7,8$) such that the variables in every factorial are non-negative,
$A,B=-(V-1)/2$ for photon subtraction, $A,B=-(V+1)/2$ for photon addition ($A$ and $B$ can take different values when different operations are applied to the two modes), and $C=\sqrt{V^2-1}/2$ for both operations.

\begin{figure}
	\centering
	\includegraphics[width=.95\linewidth]{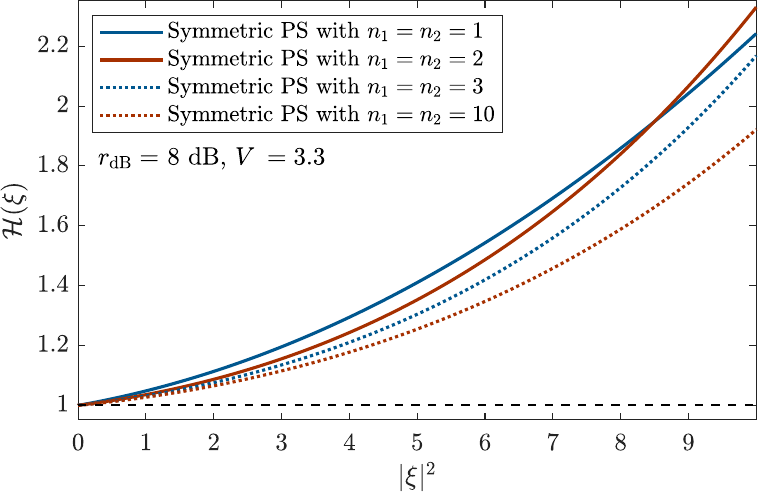}\\
	(a)\\
	\includegraphics[width=.95\linewidth]{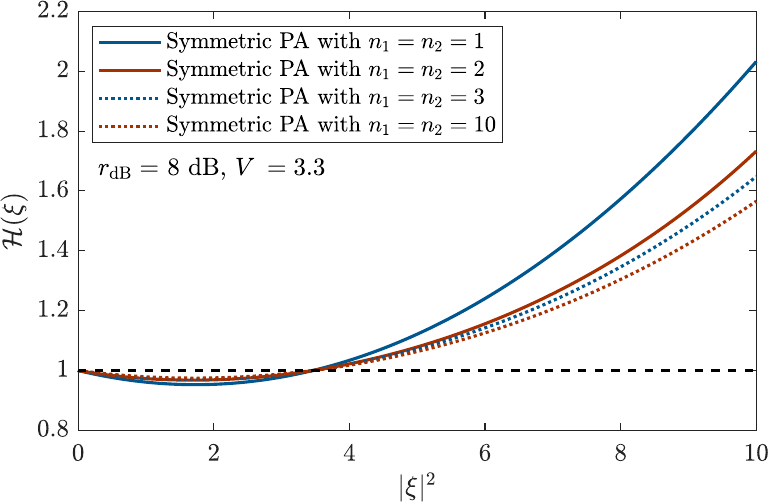}\\
	(b)\\
	\includegraphics[width=.95\linewidth]{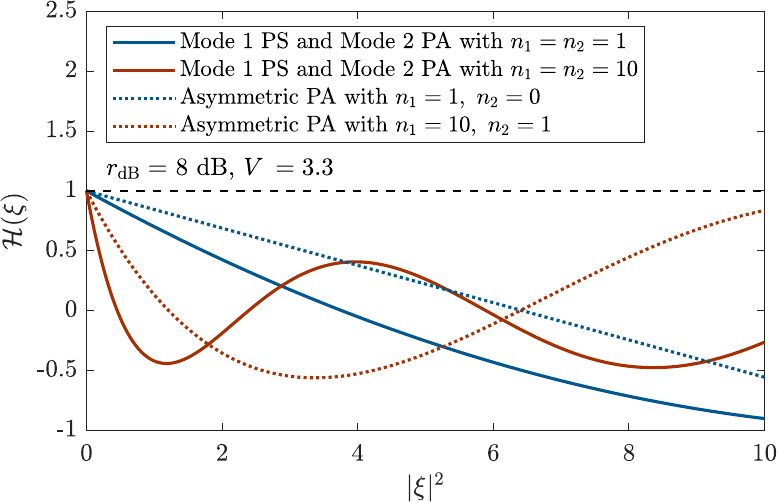}\\
	(c)
	\caption{The response ratio function $\mathcal{H}(\xi)$ for PV operations with different settings. 
	The same operation is applied to both modes of a TMSV state with 8 dB squeezing for (a) and (b). 
	In (c), different PV operations are applied to the modes of the state. (PS: photon subtraction, PA: photon addition.)}
	\label{fig:f1}
\end{figure}

Fig.~\ref{fig:f1} shows the response ratio function $\mathcal{H}(\xi)$ for PV operations with different settings.
In this paper, the squeezing of TMSV states in the dB unit is $r_\mathrm{dB}=-10\log_{10}[\exp(-2r)]$.
In \ref{fig:f1}a and \ref{fig:f1}b, we consider scenarios where both modes of a TMSV state are applied with the same operation (i.e., the symmetric scenarios with $n_1=n_2=n$).
For both PV operations the response ratios are functions of $|\xi|^2$.
For photon subtraction, we found $\mathcal{H}(\xi)>1$ for any $\xi$, showing that symmetric photon subtraction can improve teleportation fidelity for any input states.
For photon addition $\mathcal{H}(\xi)<1$ when $|\xi|^2$ is less than a certain threshold, indicating that symmetric photon addition cannot improve fidelity for input states with CF centered around the origin, e.g. coherent states and Fock states.
For input states with CF spread across the complex plain, e.g., squeezed states with high squeezing, it is possible for symmetric photon addition to improve fidelity.
For both PV operations, increasing $n$ does not increase $\mathcal{H}(\xi)$ for every $\xi$.
In \ref{fig:f1}c, we consider scenarios where different operations are applied to the modes, finding none of the asymmetric scenarios provide $\mathcal{H}(\xi)$ larger than their symmetric counterpart for any $\xi$.
Although not shown in the figures, we have also compared the response ratio function $\mathcal{H}(\xi)$ with the above PV operations for TMSV states with different squeezing, finding that the above conclusions still hold but the scales are different.

\section{CV teleportation with generalized photon varying operations}\label{sec:3}
For TMSV resource states, an upper bound to any $\mathcal{H}(\xi)$ can be written as
\begin{equation}\label{eq:hmax}
	\begin{aligned}
		\mathcal{H}_{\mathrm{max}}(\xi)&=\frac{1}{\bigchi_{\mathrm{TMSV}}(\xi, \xi^*)}\\
		&=e^{(V-\sqrt{V^2-1})|\xi|^2}.
	\end{aligned}
\end{equation}
The response ratio functions $\mathcal{H}(\xi)$ associated with PV operations $\hat{a}_{i}^{n_i}$ and $\hat{a}_i^{\dagger n_i}$ ($i=1,2$) are both polynomial functions of $|\xi|^2$, of which the degree equals $n_1+n_2$.
We have shown in the previous section that these functions do not approach $\mathcal{H}_{\mathrm{max}}(\xi)$ with increasing degree.
It remains unclear whether $\mathcal{H}(\xi)$ associated with combinations of the PV operations that approaches $\mathcal{H}_{\mathrm{max}}(\xi)$ exists.

The bound in Eq.~(\ref{eq:hmax}) can be saturated by the noiseless linear amplifier (NLA)
\begin{equation}\label{eq:nla}
	g^{\hat{a}_1^\dagger \hat{a}_1}=\sum_{n=0}^{\infty}\frac{\ln^n g}{n!}(\hat{a}_1^\dagger \hat{a}_1)^n,
\end{equation}
where the amplification gain satisfies $g\rightarrow\sqrt{(V+1)/(V-1)}$.
However, the NLA defined by Eq.~(\ref{eq:nla}) cannot be realized but approximation to the first finite terms of the series expansion of NLA is possible.
Consider the \textit{generalized} PV operations $\hat{A}_N^\dagger$ (with a conjugate transpose $\hat{A}_N$) of the form
\begin{equation}\label{eq:gePVo}
	\hat{A}_N^\dagger=\sum_{n=0}^{N}e_n\hat{a}_1^{\dagger n}\hat{a}_2^{\dagger n},
\end{equation}
where the $e_n$'s are normalized real-value coefficients.
We note that the generalized PV operation defined above does not contain any asymmetric term as we have shown in previous section that asymmetric PV operations are inferior to their symmetric counterparts.
The response ratio function for a TMSV state with the generalized PV operation given by Eq.~(\ref{eq:gePVo}) can be written as
\begin{equation}\label{eq:hgepv}
	\mathcal{H}(\xi)=\frac{1}{\bm{e}^T\bm{H}(0,0)\bm{e}}\bm{e}^T\bm{H}(\xi,\xi^*)\bm{e},
\end{equation}
where $\bm{e}=[e_0,\dots,e_N]^T$,
\begin{equation}\label{eq:hmatrix}
	\bm{H}(\xi,\xi^*)=\begin{bmatrix}
		H_{0,0,0,0}(\xi,\xi^*)&\dots&H_{N,0,N,0}(\xi,\xi^*)\\
		\vdots&\ddots&\vdots\\
		H_{0,N,0,N}(\xi,\xi^*)&\dots&H_{N,N,N,N}(\xi,\xi^*)
	\end{bmatrix},
\end{equation}
and the elements $H_{n_1,n_2,n_3,n_4}(\xi,\xi^*)$ are given by Eq.~(\ref{eq:hermite2v}) [with $A=B=-(V+1)/2$].

For given TMSV states, we propose two schemes for finding the optimal response ratio function for the generalized PV operations.
In both optimization schemes we use the integrated response ratio function as the target function to be maximized.
In the first scheme, for a fixed $N$, the optimization is taken over the vector $\bm{e}$.
The optimal $\bm{e}$ can then be written as
\begin{equation}
	\bm{e}_\mathrm{opt}=\underset{\bm{e}\in\mathbb{R}^{{N+1}}}{\mathrm{argmax}} 
	\int_{0}^{\xi_\mathrm{lim}}
	\mathcal{H}(\xi)\ d|\xi|,
\end{equation}
where we have set an upper limit $\xi_\mathrm{lim}$ for the integration with respect to $|\xi|$.

The identities
\begin{equation}\label{eq:a1a2}
	\frac{1}{\sqrt{V+1}}\hat{a}_2^\dagger\ket{\mathrm{TMSV}}=
	\frac{1}{\sqrt{V-1}}\hat{a}_1\ket{\mathrm{TMSV}},
\end{equation}
and \cite{blasiak2007combinatorics}
\begin{equation}\label{eq:a1a1n}
	(\hat{a}_1^\dagger \hat{a}_1)^n=\sum_{m=1}^{n}	
	\frac{1}{m!}
	\sum_{j=1}^{m}
	\begin{pmatrix}
		m\\j
	\end{pmatrix}
	(-1)^{m-j}
	j^n
	(\hat{a}_1^\dagger)^m\hat{a}_1^m,
\end{equation}
show that any polynomials of $({\hat{a}_1}^\dagger \hat{a}_1)$ with degree $N$ can be map to an equivalent operator $\hat{A}_N^\dagger$ (or $\hat{A}_N$) when applied to TMSV states.
As a result, the generalized PV operations for TMSV states can be realized by a cascaded application of a coherent PV operation \cite{chatterjee2012Nonclassical} or a parallel application of photon catalysis \cite{2021_354} on one mode of the state.
In the second optimization scheme, 
using Eqs.~(\ref{eq:a1a2}) and (\ref{eq:a1a1n}),
we first set $\bm{e}$ such that the operator $\hat{A}_N^\dagger$ equals the first $N$ terms of the NLA defined by Eq.~(\ref{eq:nla}), expressing $\bm{e}$ as a function of $g$.
We then optimize the integrated response ratio function over $g$.

\begin{figure}
	\centering
	\includegraphics[width=.95\linewidth]{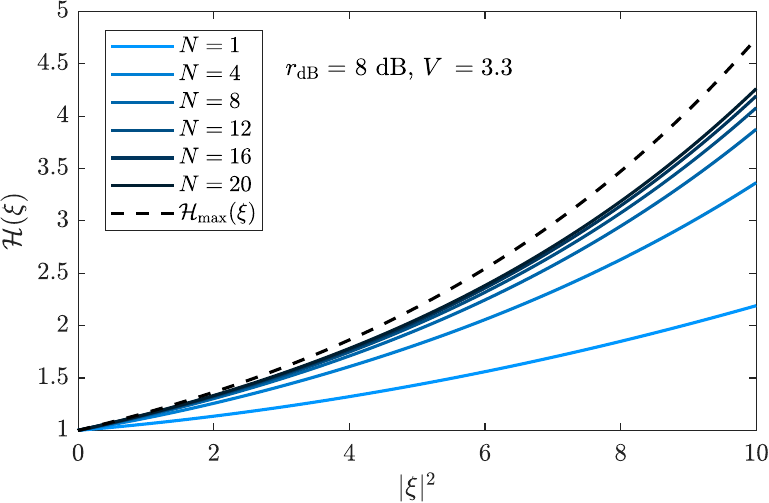}\\
	(a)\\
	\includegraphics[width=.95\linewidth]{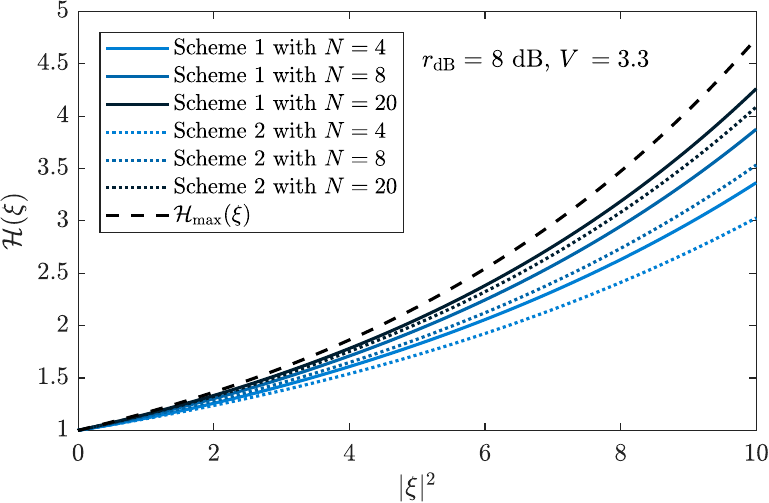}\\
	(b)
	\includegraphics[width=.95\linewidth]{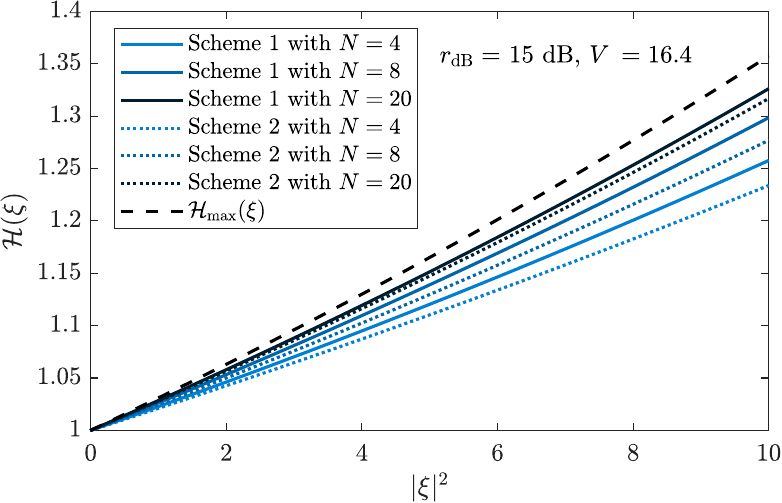}\\
	(c)
	\caption{The response ratio function $\mathcal{H}(\xi)$ for TMSV states with generalized PV operations with different parameter vector $\bm{e}$. 
		In (a) $\bm{e}$ is set such that the integrated response ratio function is maximized.
		In (b) and (c) the two schemes are compared.
		TMSV states with higher initial squeezing are considered in (c).
		The upper bound to the response ratio function $\mathcal{H}_{\mathrm{max}}(\xi)$ is defined in Eq.~(\ref{eq:hmax}).
		(Scheme 1: optimization taken over $\bm{e}$. Scheme 2: optimization take over $g$.)}
	\label{fig:f2}
\end{figure}

Fig.~\ref{fig:f2} shows the response ratio function $\mathcal{H}(\xi)$ for the generalized PV operations $\hat{A}_N^\dagger$ with optimized parameters.
In this paper, we adopt a particle swarm algorithm for all optimization problems.
In \ref{fig:f2}a the parameter vector $\bm{e}$ is set to maximize the integrated response ratio function (with $\xi_\mathrm{lim}=2$).
Distinct from the PV operations discussed in the previous section, the generalized PV operations always provide increasing $\mathcal{H}(\xi)$ with increasing $N$.
The response ratio function also approaches the upper bound $\mathcal{H}_{\mathrm{max}}(\xi)$ as $N$ grows.
However, we are unable to show that $\mathcal{H}_{\mathrm{max}}(\xi)$ is saturated when $N$ approaches infinite because we cannot calculate $\bm{H}(\xi,\xi^*)$ accurately with $N>20$.
Fig.~\ref{fig:f2}b and \ref{fig:f2}c compare the two optimization schemes.
It is not surprising that the optimization over $\bm{e}$ outperforms the optimization over $g$ because the former has a higher degree of freedom in optimization.


\section{Teleportation with other resource states and photon varying operations}\label{sec:4}
In this section, we extend our studies to some generalized forms of TMSV states.
\subsection{Two-mode squeezed coherent states}
Two-mode squeezed coherent (TMSC) states have drawn research attention in recent years due to their ability to improve measurement-independent quantum key distribution \cite{kumar2019coherence,singh2021non,2023_473}.
A general TMSC state can be defined by \cite{selvadoray1997phase}
\begin{equation}\label{eq:tmsc}
	\ket{\mathrm{TMSC}}=\hat{D}_1(z_1)\hat{D}_2(z_2)\hat{S}(r)\ket{0,0}_{1,2},
\end{equation}
where $\hat{D}_1(z_1),\ \hat{D}_2(z_2)$ are displacement operators with $z_1,\ z_2\in\mathbb{C}$.
The above state has an alternative definition, which is given by swapping the order of the squeezing operator and the displacement operators,
\begin{equation}
	\ket{\mathrm{TMSC}}=\hat{S}(r)\hat{D}_1(\tilde{z}_1)\hat{D}_2(\tilde{z}_2)\ket{0,0}_{1,2},
\end{equation}
where $\tilde{z}_1=z_1\cosh{r}-z_2^*\sinh{r}$ and $\tilde{z}_2=z_2\cosh{r}-z_1^*\sinh{r}$
In this work, we will use the definition in Eq.~(\ref{eq:tmsc}). 
The CF for a TMSC state can then be written as
\begin{equation}
	\bigchi_{\mathrm{TMSC}}(\xi_1,\xi_2)=e^{\xi_1z_1^*-\xi_1^*z_1}
	e^{\xi_2z_2^*-\xi_2^*z_2}
	\bigchi_{\mathrm{TMSV}}(\xi_1,\xi_2),
\end{equation}
where $\bigchi_{\mathrm{TMSV}}(\xi_1,\xi_2)$ is the CF for the TMSV state in Eq.~(\ref{eq:cfTMSV}).

\begin{figure}
	\centering
	\includegraphics[width=0.8\linewidth]{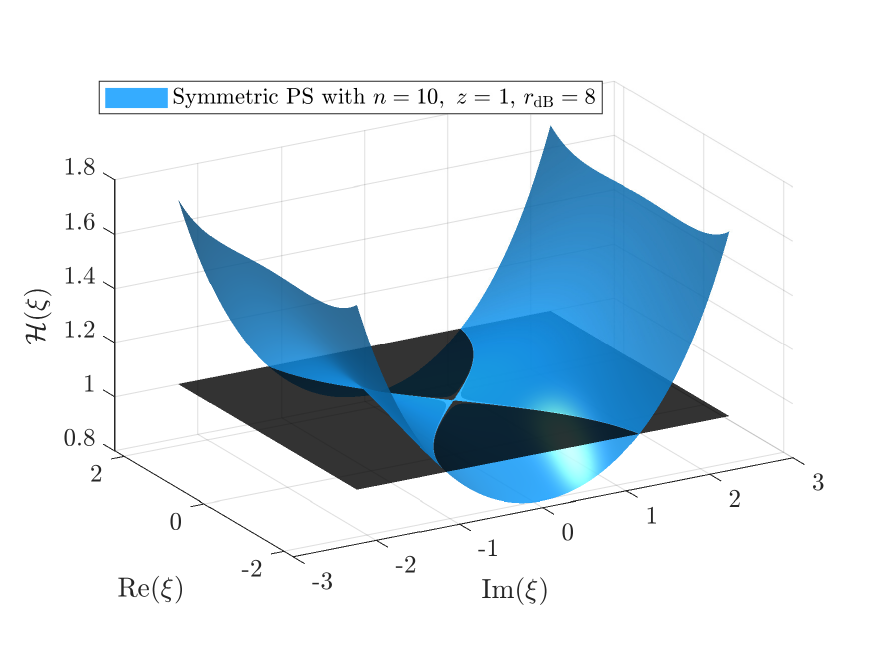}\\(a)\\
	\includegraphics[width=0.8\linewidth]{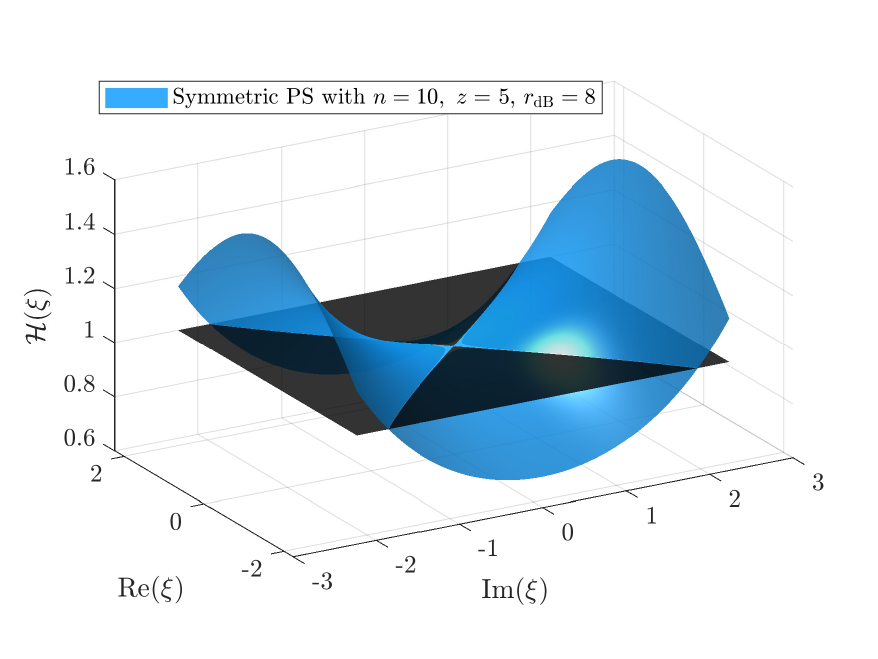}\\(b)\\
	\caption{The response ratio functions for two-mode squeezed coherent states with PV operations are shown in (a) and (b) with different displacement $z$. Both functions are symmetric with respect to the imaginary axis $\mathrm{Im}(\xi)$.}
	\label{fig:f3d}
\end{figure}

Consider the case for $z_1=z_2=z\in\mathbb{R}$, the response function for a TMSC state equals that of a TMSV state, $\bigchi_{\mathrm{TMSV}}(\xi,\xi^*)$.
Similar to before, the response ratio functions for TMSC states with PV operations can be calculated using Eq.~(\ref{eq:pvcf}), which are not shown here for conciseness.
Examples of such functions are shown in Fig.~\ref{fig:f3d}a and \ref{fig:f3d}b, where we have adopted a 3D plot as they cannot be expressed as a function of $|\xi|$.
Fig.~\ref{fig:f3d} shows that for a given PV operations $\mathcal{H}(\xi)$ decreases as $z$ increases.

\begin{figure}
	\centering
	\includegraphics[width=.95\linewidth]{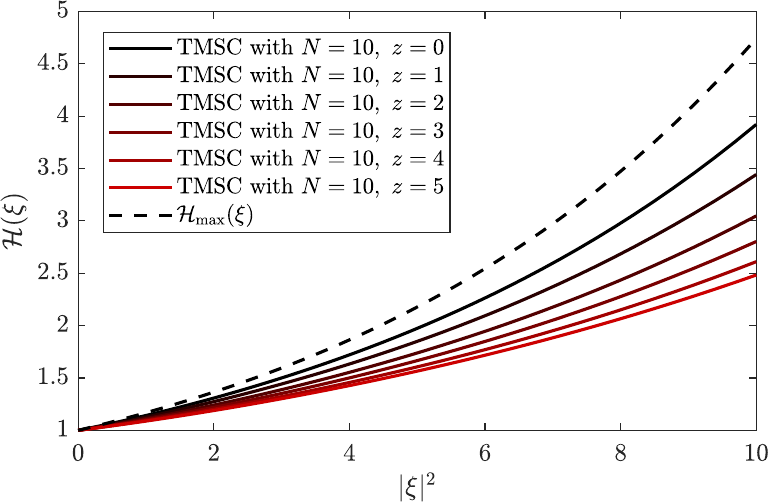}\\(a)\\
	\includegraphics[width=.95\linewidth]{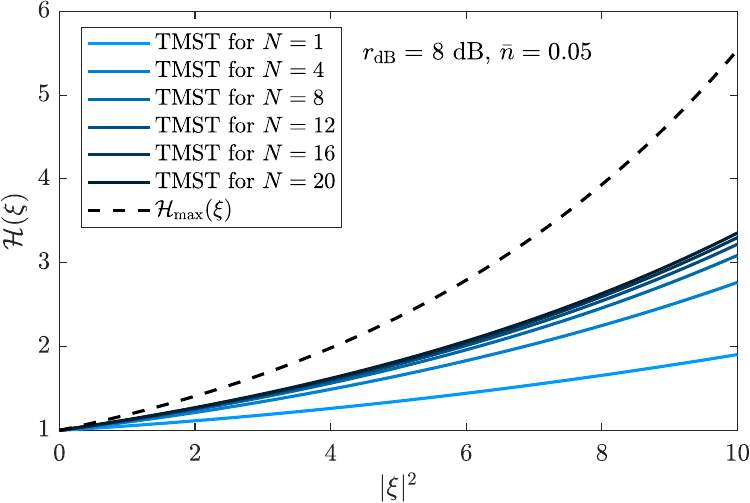}\\(b)\\
	\caption{The response ratio function $\mathcal{H}(\xi)$  with generalized PV operations $\hat{A}_N$ and optimized $\bm{e}$ for 
		(a) two-mode squeezed coherent states 
		(b) two-mode squeezed thermal states. For all figures the initial squeezing of the resource state is set to 8 dB.}
	\label{fig:f4}
\end{figure}

Next, we consider TMSC states with generalized PV operations $\hat{A}_N$ and $\hat{A}_N^\dagger$.
Recall from Eq.~(\ref{eq:hmax}) the upper bound to states with the response function $\bigchi_{\mathrm{TMSV}}(\xi,\xi^*)$ is a polynomials of $|\xi|^2$.
For $\hat{A}_N$ and $\hat{A}^\dagger_N$ the non-$|\xi|^2$ terms in $\mathcal{H}(\xi)$ can be eliminated by a proper setting of the parameter vector $\bm{e}$.
Fig.~\ref{fig:f4}a shows the optimized response ratio function for the TMSC states with different displacement $z$, where $\mathcal{H}(\xi)$ is optimized over $\bm{e}$ independently for each state.
Similar to the results in Fig.~\ref{fig:f3d}, the $\mathcal{H}(\xi)$ for TMSC states with the generalized PV operations also decreases with increasing $z$.
The results shown suggest that to obtain the maximal improvement offered by the PV operations, the mean values of TMSC states must be displaced to zero before the operations.

\subsection{Two-mode squeezed thermal states}
Two-mode squeezed thermal (TMST) states are another generalization of the TMSV states.
A TMST state can be defined by
\begin{equation}
	\hat{\rho}_{\mathrm{TMST}}=\hat{S}(r)\hat{\rho}_\mathrm{th1}
	\hat{\rho}_\mathrm{th2}
	\hat{S}^\dagger(r),
\end{equation}
where
\begin{equation}
	\hat{\rho}_{\mathrm{th}i}=\sum_{n=0}^{\infty}\frac{\bar{n}_i^2}{(\bar{n}_i+1)^{n+1}}
	\ket{n}_i\bra{n}_i,
\end{equation}
$i=1,2$ and the $\hat{\rho}_{\mathrm{th}i}$'s are thermal states with mean photon number $\bar{n}_i$.
Consider the case for $\bar{n}_1=\bar{n}_2=\bar{n}$, the CF for the TMST state can be written as
\begin{equation}
	\begin{aligned}			
	&\bigchi_{\mathrm{TMST}}(\xi_1,\xi_2)\\
	&\hskip 10pt=e^{-\frac{1}{2}\left[V\left(|\xi_1|^2+\xi_2|^2\right)-\sqrt{V^2-1}
		\left(\xi_1\xi_2+{\xi_1}^*{\xi_2}^*\right)\right](2\bar{n}+1)}.
	\end{aligned}
\end{equation}
The upper bound to the response ration function for a TMST state is $\mathcal{H}_{\mathrm{max}}(\xi)=e^{(2\bar{n}+1)(V-\sqrt{V^2-1})|\xi|^2}$.
Fig.~\ref{fig:f4}b 
shows the optimized response ratio function for the TMST states with $\hat{A}_N$, where again $\mathcal{H}(\xi)$ is optimized over $\bm{e}$.
As $N$ increases the $\mathcal{H}(\xi)$ converges to a curve below $\mathcal{H}_{\mathrm{max}}(\xi)$, showing that the generalized PV operations cannot reduce noise in the resource state.
Although not displayed in the figures, for $\hat{A}_N^\dagger$ the optimized $\mathcal{H}(\xi)$ converges to a curve above that for $\hat{A}_N$ because the relation in Eq.~(\ref{eq:a1a2}) is not hold for TMST states.

\subsection{Dissipative TMSV states}
The distribution of entangled states is a prerequisite for quantum teleportation.
For TMSV states passed through dissipative environments, the generalized PV operations cannot be realized in a local fashion.
Therefore, we only consider the case where the PV operations are applied before the channel transmission of the TMSV state.
Consider two independent lossy channels modeled by beam splitters with transmissivity $T_\mathrm{1}$ and $T_\mathrm{2}$.
A TMSV state is first applied with a generalized PV operation, of which the two modes are then transmitted through the channels.
We define a new function $\mathcal{H}'(\xi)$ as the ratio between the response function for a TMSV state first applied with a generalized PV operation and then distributed over the channels and the response function for a TMSV state directly distributed over the channels.
We find that $\mathcal{H}'(\xi)>1$ is always satisfied for $T_\mathrm{1}=T_\mathrm{2}$. 
For $T_\mathrm{1}\neq T_\mathrm{2}$, $\mathcal{H}'(\xi)$ might drop below unit for certain area of $\xi$.
The above results coincide with the conclusion in \cite{mivcuda2012noiseless} that TMSV states with higher squeezing are more sensitive to channels with asymmetric losses.

\section{Conclusion}\label{sec:5}
In this paper, we built a framework for PV operations in the multi-mode setting, expressing the characteristic function of the photon-varied two-mode Gaussian state as a multi-variable multi-index Hermite-Gaussian function.
Based on the framework built, we introduced the response ratio function, which is a performance metric suitable for arbitrary input states.
For a given teleportation resource state, the response ratio function associated with the PV operations can be easily transformed to the fidelity for any input states.
We then proposed generalized PV operations and investigated the use of such operations in various Gaussian states.
Our results show that the generalized PV operations that approach the noiseless linear amplification provide the most improvement in teleportation with TMSV states.

\section*{Acknowledgment}
The authors acknowledge financial support from China Postdoctoral Science Foundation (Fund No. 2023M741315) and Central China Normal University (Fund No. CCNU23XJ014).

\bibliography{mybib}

\begin{thebibliography}{69}%
\makeatletter
\providecommand \@ifxundefined [1]{%
 \@ifx{#1\undefined}
}%
\providecommand \@ifnum [1]{%
 \ifnum #1\expandafter \@firstoftwo
 \else \expandafter \@secondoftwo
 \fi
}%
\providecommand \@ifx [1]{%
 \ifx #1\expandafter \@firstoftwo
 \else \expandafter \@secondoftwo
 \fi
}%
\providecommand \natexlab [1]{#1}%
\providecommand \enquote  [1]{``#1''}%
\providecommand \bibnamefont  [1]{#1}%
\providecommand \bibfnamefont [1]{#1}%
\providecommand \citenamefont [1]{#1}%
\providecommand \href@noop [0]{\@secondoftwo}%
\providecommand \href [0]{\begingroup \@sanitize@url \@href}%
\providecommand \@href[1]{\@@startlink{#1}\@@href}%
\providecommand \@@href[1]{\endgroup#1\@@endlink}%
\providecommand \@sanitize@url [0]{\catcode `\\12\catcode `\$12\catcode
  `\&12\catcode `\#12\catcode `\^12\catcode `\_12\catcode `\%12\relax}%
\providecommand \@@startlink[1]{}%
\providecommand \@@endlink[0]{}%
\providecommand \url  [0]{\begingroup\@sanitize@url \@url }%
\providecommand \@url [1]{\endgroup\@href {#1}{\urlprefix }}%
\providecommand \urlprefix  [0]{URL }%
\providecommand \Eprint [0]{\href }%
\providecommand \doibase [0]{https://doi.org/}%
\providecommand \selectlanguage [0]{\@gobble}%
\providecommand \bibinfo  [0]{\@secondoftwo}%
\providecommand \bibfield  [0]{\@secondoftwo}%
\providecommand \translation [1]{[#1]}%
\providecommand \BibitemOpen [0]{}%
\providecommand \bibitemStop [0]{}%
\providecommand \bibitemNoStop [0]{.\EOS\space}%
\providecommand \EOS [0]{\spacefactor3000\relax}%
\providecommand \BibitemShut  [1]{\csname bibitem#1\endcsname}%
\let\auto@bib@innerbib\@empty
\bibitem [{\citenamefont {Yonezawa}\ \emph {et~al.}(2004)\citenamefont
  {Yonezawa}, \citenamefont {Aoki},\ and\ \citenamefont
  {Furusawa}}]{yonezawa2004demonstration}%
  \BibitemOpen
  \bibfield  {author} {\bibinfo {author} {\bibfnamefont {H.}~\bibnamefont
  {Yonezawa}}, \bibinfo {author} {\bibfnamefont {T.}~\bibnamefont {Aoki}},\
  and\ \bibinfo {author} {\bibfnamefont {A.}~\bibnamefont {Furusawa}},\
  }\bibfield  {title} {\bibinfo {title} {Demonstration of a quantum
  teleportation network for continuous variables},\ }\href@noop {} {\bibfield
  {journal} {\bibinfo  {journal} {Nature}\ }\textbf {\bibinfo {volume} {431}},\
  \bibinfo {pages} {430} (\bibinfo {year} {2004})}\BibitemShut {NoStop}%
\bibitem [{\citenamefont {Jin}\ \emph {et~al.}(2010)\citenamefont {Jin},
  \citenamefont {Ren}, \citenamefont {Yang}, \citenamefont {Yi}, \citenamefont
  {Zhou}, \citenamefont {Xu}, \citenamefont {Wang}, \citenamefont {Yang},
  \citenamefont {Hu}, \citenamefont {Jiang} \emph
  {et~al.}}]{jin2010experimental}%
  \BibitemOpen
  \bibfield  {author} {\bibinfo {author} {\bibfnamefont {X.-M.}\ \bibnamefont
  {Jin}}, \bibinfo {author} {\bibfnamefont {J.-G.}\ \bibnamefont {Ren}},
  \bibinfo {author} {\bibfnamefont {B.}~\bibnamefont {Yang}}, \bibinfo {author}
  {\bibfnamefont {Z.-H.}\ \bibnamefont {Yi}}, \bibinfo {author} {\bibfnamefont
  {F.}~\bibnamefont {Zhou}}, \bibinfo {author} {\bibfnamefont {X.-F.}\
  \bibnamefont {Xu}}, \bibinfo {author} {\bibfnamefont {S.-K.}\ \bibnamefont
  {Wang}}, \bibinfo {author} {\bibfnamefont {D.}~\bibnamefont {Yang}}, \bibinfo
  {author} {\bibfnamefont {Y.-F.}\ \bibnamefont {Hu}}, \bibinfo {author}
  {\bibfnamefont {S.}~\bibnamefont {Jiang}}, \emph {et~al.},\ }\bibfield
  {title} {\bibinfo {title} {Experimental free-space quantum teleportation},\
  }\href@noop {} {\bibfield  {journal} {\bibinfo  {journal} {Nature photonics}\
  }\textbf {\bibinfo {volume} {4}},\ \bibinfo {pages} {376} (\bibinfo {year}
  {2010})}\BibitemShut {NoStop}%
\bibitem [{\citenamefont {Ma}\ \emph {et~al.}(2012)\citenamefont {Ma},
  \citenamefont {Herbst}, \citenamefont {Scheidl}, \citenamefont {Wang},
  \citenamefont {Kropatschek}, \citenamefont {Naylor}, \citenamefont
  {Wittmann}, \citenamefont {Mech}, \citenamefont {Kofler}, \citenamefont
  {Anisimova} \emph {et~al.}}]{ma2012quantum}%
  \BibitemOpen
  \bibfield  {author} {\bibinfo {author} {\bibfnamefont {X.-S.}\ \bibnamefont
  {Ma}}, \bibinfo {author} {\bibfnamefont {T.}~\bibnamefont {Herbst}}, \bibinfo
  {author} {\bibfnamefont {T.}~\bibnamefont {Scheidl}}, \bibinfo {author}
  {\bibfnamefont {D.}~\bibnamefont {Wang}}, \bibinfo {author} {\bibfnamefont
  {S.}~\bibnamefont {Kropatschek}}, \bibinfo {author} {\bibfnamefont
  {W.}~\bibnamefont {Naylor}}, \bibinfo {author} {\bibfnamefont
  {B.}~\bibnamefont {Wittmann}}, \bibinfo {author} {\bibfnamefont
  {A.}~\bibnamefont {Mech}}, \bibinfo {author} {\bibfnamefont {J.}~\bibnamefont
  {Kofler}}, \bibinfo {author} {\bibfnamefont {E.}~\bibnamefont {Anisimova}},
  \emph {et~al.},\ }\bibfield  {title} {\bibinfo {title} {Quantum teleportation
  over 143 kilometres using active feed-forward},\ }\href@noop {} {\bibfield
  {journal} {\bibinfo  {journal} {Nature}\ }\textbf {\bibinfo {volume} {489}},\
  \bibinfo {pages} {269} (\bibinfo {year} {2012})}\BibitemShut {NoStop}%
\bibitem [{\citenamefont {Ren}\ \emph {et~al.}(2017)\citenamefont {Ren},
  \citenamefont {Xu}, \citenamefont {Yong}, \citenamefont {Zhang},
  \citenamefont {Liao}, \citenamefont {Yin}, \citenamefont {Liu}, \citenamefont
  {Cai}, \citenamefont {Yang}, \citenamefont {Li} \emph
  {et~al.}}]{ren2017ground}%
  \BibitemOpen
  \bibfield  {author} {\bibinfo {author} {\bibfnamefont {J.-G.}\ \bibnamefont
  {Ren}}, \bibinfo {author} {\bibfnamefont {P.}~\bibnamefont {Xu}}, \bibinfo
  {author} {\bibfnamefont {H.-L.}\ \bibnamefont {Yong}}, \bibinfo {author}
  {\bibfnamefont {L.}~\bibnamefont {Zhang}}, \bibinfo {author} {\bibfnamefont
  {S.-K.}\ \bibnamefont {Liao}}, \bibinfo {author} {\bibfnamefont
  {J.}~\bibnamefont {Yin}}, \bibinfo {author} {\bibfnamefont {W.-Y.}\
  \bibnamefont {Liu}}, \bibinfo {author} {\bibfnamefont {W.-Q.}\ \bibnamefont
  {Cai}}, \bibinfo {author} {\bibfnamefont {M.}~\bibnamefont {Yang}}, \bibinfo
  {author} {\bibfnamefont {L.}~\bibnamefont {Li}}, \emph {et~al.},\ }\bibfield
  {title} {\bibinfo {title} {Ground-to-satellite quantum teleportation},\
  }\href@noop {} {\bibfield  {journal} {\bibinfo  {journal} {Nature}\ }\textbf
  {\bibinfo {volume} {549}},\ \bibinfo {pages} {70} (\bibinfo {year}
  {2017})}\BibitemShut {NoStop}%
\bibitem [{\citenamefont {Lago-Rivera}\ \emph {et~al.}(2023)\citenamefont
  {Lago-Rivera}, \citenamefont {Rakonjac}, \citenamefont {Grandi},\ and\
  \citenamefont {Riedmatten}}]{lago2023long}%
  \BibitemOpen
  \bibfield  {author} {\bibinfo {author} {\bibfnamefont {D.}~\bibnamefont
  {Lago-Rivera}}, \bibinfo {author} {\bibfnamefont {J.~V.}\ \bibnamefont
  {Rakonjac}}, \bibinfo {author} {\bibfnamefont {S.}~\bibnamefont {Grandi}},\
  and\ \bibinfo {author} {\bibfnamefont {H.~d.}\ \bibnamefont {Riedmatten}},\
  }\bibfield  {title} {\bibinfo {title} {Long distance multiplexed quantum
  teleportation from a telecom photon to a solid-state qubit},\ }\href@noop {}
  {\bibfield  {journal} {\bibinfo  {journal} {Nature Communications}\ }\textbf
  {\bibinfo {volume} {14}},\ \bibinfo {pages} {1889} (\bibinfo {year}
  {2023})}\BibitemShut {NoStop}%
\bibitem [{\citenamefont {Zhao}\ \emph {et~al.}(2023)\citenamefont {Zhao},
  \citenamefont {Jeng}, \citenamefont {Conlon}, \citenamefont {Tserkis},
  \citenamefont {Shajilal}, \citenamefont {Liu}, \citenamefont {Ralph},
  \citenamefont {Assad},\ and\ \citenamefont {Lam}}]{zhao2023enhancing}%
  \BibitemOpen
  \bibfield  {author} {\bibinfo {author} {\bibfnamefont {J.}~\bibnamefont
  {Zhao}}, \bibinfo {author} {\bibfnamefont {H.}~\bibnamefont {Jeng}}, \bibinfo
  {author} {\bibfnamefont {L.~O.}\ \bibnamefont {Conlon}}, \bibinfo {author}
  {\bibfnamefont {S.}~\bibnamefont {Tserkis}}, \bibinfo {author} {\bibfnamefont
  {B.}~\bibnamefont {Shajilal}}, \bibinfo {author} {\bibfnamefont
  {K.}~\bibnamefont {Liu}}, \bibinfo {author} {\bibfnamefont {T.~C.}\
  \bibnamefont {Ralph}}, \bibinfo {author} {\bibfnamefont {S.~M.}\ \bibnamefont
  {Assad}},\ and\ \bibinfo {author} {\bibfnamefont {P.~K.}\ \bibnamefont
  {Lam}},\ }\bibfield  {title} {\bibinfo {title} {Enhancing quantum
  teleportation efficacy with noiseless linear amplification},\ }\href@noop {}
  {\bibfield  {journal} {\bibinfo  {journal} {Nature Communications}\ }\textbf
  {\bibinfo {volume} {14}},\ \bibinfo {pages} {4745} (\bibinfo {year}
  {2023})}\BibitemShut {NoStop}%
\bibitem [{\citenamefont {Vahlbruch}\ \emph {et~al.}(2016)\citenamefont
  {Vahlbruch}, \citenamefont {Mehmet}, \citenamefont {Danzmann},\ and\
  \citenamefont {Schnabel}}]{vahlbruch2016detection}%
  \BibitemOpen
  \bibfield  {author} {\bibinfo {author} {\bibfnamefont {H.}~\bibnamefont
  {Vahlbruch}}, \bibinfo {author} {\bibfnamefont {M.}~\bibnamefont {Mehmet}},
  \bibinfo {author} {\bibfnamefont {K.}~\bibnamefont {Danzmann}},\ and\
  \bibinfo {author} {\bibfnamefont {R.}~\bibnamefont {Schnabel}},\ }\bibfield
  {title} {\bibinfo {title} {Detection of 15 {dB} squeezed states of light and
  their application for the absolute calibration of photoelectric quantum
  efficiency},\ }\href@noop {} {\bibfield  {journal} {\bibinfo  {journal}
  {Physical Review Letters}\ }\textbf {\bibinfo {volume} {117}},\ \bibinfo
  {pages} {110801} (\bibinfo {year} {2016})}\BibitemShut {NoStop}%
\bibitem [{\citenamefont {Mi{\v{c}}uda}\ \emph {et~al.}(2012)\citenamefont
  {Mi{\v{c}}uda}, \citenamefont {Straka}, \citenamefont {Mikov{\'a}},
  \citenamefont {Du{\v{s}}ek}, \citenamefont {Cerf}, \citenamefont
  {Fiur{\'a}{\v{s}}ek},\ and\ \citenamefont
  {Je{\v{z}}ek}}]{mivcuda2012noiseless}%
  \BibitemOpen
  \bibfield  {author} {\bibinfo {author} {\bibfnamefont {M.}~\bibnamefont
  {Mi{\v{c}}uda}}, \bibinfo {author} {\bibfnamefont {I.}~\bibnamefont
  {Straka}}, \bibinfo {author} {\bibfnamefont {M.}~\bibnamefont {Mikov{\'a}}},
  \bibinfo {author} {\bibfnamefont {M.}~\bibnamefont {Du{\v{s}}ek}}, \bibinfo
  {author} {\bibfnamefont {N.~J.}\ \bibnamefont {Cerf}}, \bibinfo {author}
  {\bibfnamefont {J.}~\bibnamefont {Fiur{\'a}{\v{s}}ek}},\ and\ \bibinfo
  {author} {\bibfnamefont {M.}~\bibnamefont {Je{\v{z}}ek}},\ }\bibfield
  {title} {\bibinfo {title} {Noiseless loss suppression in quantum optical
  communication},\ }\href@noop {} {\bibfield  {journal} {\bibinfo  {journal}
  {Physical Review Letters}\ }\textbf {\bibinfo {volume} {109}},\ \bibinfo
  {pages} {180503} (\bibinfo {year} {2012})}\BibitemShut {NoStop}%
\bibitem [{\citenamefont {Huang}\ \emph {et~al.}(2013)\citenamefont {Huang},
  \citenamefont {He}, \citenamefont {Fang},\ and\ \citenamefont
  {Zeng}}]{huang2013performance}%
  \BibitemOpen
  \bibfield  {author} {\bibinfo {author} {\bibfnamefont {P.}~\bibnamefont
  {Huang}}, \bibinfo {author} {\bibfnamefont {G.}~\bibnamefont {He}}, \bibinfo
  {author} {\bibfnamefont {J.}~\bibnamefont {Fang}},\ and\ \bibinfo {author}
  {\bibfnamefont {G.}~\bibnamefont {Zeng}},\ }\bibfield  {title} {\bibinfo
  {title} {Performance improvement of continuous-variable quantum key
  distribution via photon subtraction},\ }\href@noop {} {\bibfield  {journal}
  {\bibinfo  {journal} {Physical Review A}\ }\textbf {\bibinfo {volume} {87}},\
  \bibinfo {pages} {012317} (\bibinfo {year} {2013})}\BibitemShut {NoStop}%
\bibitem [{\citenamefont {Borelli}\ \emph {et~al.}(2016)\citenamefont
  {Borelli}, \citenamefont {Aguiar}, \citenamefont {Roversi},\ and\
  \citenamefont {Vidiellabarranco}}]{borelli2016quantum}%
  \BibitemOpen
  \bibfield  {author} {\bibinfo {author} {\bibfnamefont {L.~F.}\ \bibnamefont
  {Borelli}}, \bibinfo {author} {\bibfnamefont {L.~S.}\ \bibnamefont {Aguiar}},
  \bibinfo {author} {\bibfnamefont {J.~A.}\ \bibnamefont {Roversi}},\ and\
  \bibinfo {author} {\bibfnamefont {A.}~\bibnamefont {Vidiellabarranco}},\
  }\bibfield  {title} {\bibinfo {title} {Quantum key distribution using
  continuous-variable non-{G}aussian states},\ }\href@noop {} {\bibfield
  {journal} {\bibinfo  {journal} {Quantum Information Processing}\ }\textbf
  {\bibinfo {volume} {15}},\ \bibinfo {pages} {893} (\bibinfo {year}
  {2016})}\BibitemShut {NoStop}%
\bibitem [{\citenamefont {Li}\ \emph {et~al.}(2016)\citenamefont {Li},
  \citenamefont {Zhang}, \citenamefont {Wang}, \citenamefont {Xu},
  \citenamefont {Peng},\ and\ \citenamefont {Guo}}]{li2016non-gaussian}%
  \BibitemOpen
  \bibfield  {author} {\bibinfo {author} {\bibfnamefont {Z.}~\bibnamefont
  {Li}}, \bibinfo {author} {\bibfnamefont {Y.}~\bibnamefont {Zhang}}, \bibinfo
  {author} {\bibfnamefont {X.}~\bibnamefont {Wang}}, \bibinfo {author}
  {\bibfnamefont {B.}~\bibnamefont {Xu}}, \bibinfo {author} {\bibfnamefont
  {X.}~\bibnamefont {Peng}},\ and\ \bibinfo {author} {\bibfnamefont
  {H.}~\bibnamefont {Guo}},\ }\bibfield  {title} {\bibinfo {title}
  {Non-{G}aussian postselection and virtual photon subtraction in
  continuous-variable quantum key distribution},\ }\href@noop {} {\bibfield
  {journal} {\bibinfo  {journal} {Physical Review A}\ }\textbf {\bibinfo
  {volume} {93}},\ \bibinfo {pages} {012310} (\bibinfo {year}
  {2016})}\BibitemShut {NoStop}%
\bibitem [{\citenamefont {Guo}\ \emph {et~al.}(2017)\citenamefont {Guo},
  \citenamefont {Liao}, \citenamefont {Wang}, \citenamefont {Huang},
  \citenamefont {Huang},\ and\ \citenamefont {Zeng}}]{guo2017performance}%
  \BibitemOpen
  \bibfield  {author} {\bibinfo {author} {\bibfnamefont {Y.}~\bibnamefont
  {Guo}}, \bibinfo {author} {\bibfnamefont {Q.}~\bibnamefont {Liao}}, \bibinfo
  {author} {\bibfnamefont {Y.}~\bibnamefont {Wang}}, \bibinfo {author}
  {\bibfnamefont {D.}~\bibnamefont {Huang}}, \bibinfo {author} {\bibfnamefont
  {P.}~\bibnamefont {Huang}},\ and\ \bibinfo {author} {\bibfnamefont
  {G.}~\bibnamefont {Zeng}},\ }\bibfield  {title} {\bibinfo {title}
  {Performance improvement of continuous-variable quantum key distribution with
  an entangled source in the middle via photon subtraction},\ }\href@noop {}
  {\bibfield  {journal} {\bibinfo  {journal} {Physical Review A}\ }\textbf
  {\bibinfo {volume} {95}} (\bibinfo {year} {2017})}\BibitemShut {NoStop}%
\bibitem [{\citenamefont {Zhao}\ \emph {et~al.}(2017)\citenamefont {Zhao},
  \citenamefont {Zhang}, \citenamefont {Li}, \citenamefont {Yu},\ and\
  \citenamefont {Guo}}]{zhao2017improvement}%
  \BibitemOpen
  \bibfield  {author} {\bibinfo {author} {\bibfnamefont {Y.}~\bibnamefont
  {Zhao}}, \bibinfo {author} {\bibfnamefont {Y.}~\bibnamefont {Zhang}},
  \bibinfo {author} {\bibfnamefont {Z.}~\bibnamefont {Li}}, \bibinfo {author}
  {\bibfnamefont {S.}~\bibnamefont {Yu}},\ and\ \bibinfo {author}
  {\bibfnamefont {H.}~\bibnamefont {Guo}},\ }\bibfield  {title} {\bibinfo
  {title} {Improvement of two-way continuous-variable quantum key distribution
  with virtual photon subtraction},\ }\href@noop {} {\bibfield  {journal}
  {\bibinfo  {journal} {Quantum Information Processing}\ }\textbf {\bibinfo
  {volume} {16}},\ \bibinfo {pages} {184} (\bibinfo {year} {2017})}\BibitemShut
  {NoStop}%
\bibitem [{\citenamefont {Ma}\ \emph {et~al.}(2018)\citenamefont {Ma},
  \citenamefont {Huang}, \citenamefont {Bai}, \citenamefont {Wang},
  \citenamefont {Bao},\ and\ \citenamefont {Zeng}}]{ma2018continuous-variable}%
  \BibitemOpen
  \bibfield  {author} {\bibinfo {author} {\bibfnamefont {H.}~\bibnamefont
  {Ma}}, \bibinfo {author} {\bibfnamefont {P.}~\bibnamefont {Huang}}, \bibinfo
  {author} {\bibfnamefont {D.}~\bibnamefont {Bai}}, \bibinfo {author}
  {\bibfnamefont {S.}~\bibnamefont {Wang}}, \bibinfo {author} {\bibfnamefont
  {W.}~\bibnamefont {Bao}},\ and\ \bibinfo {author} {\bibfnamefont
  {G.}~\bibnamefont {Zeng}},\ }\bibfield  {title} {\bibinfo {title}
  {Continuous-variable measurement-device-independent quantum key distribution
  with photon subtraction},\ }\href@noop {} {\bibfield  {journal} {\bibinfo
  {journal} {Physical Review A}\ }\textbf {\bibinfo {volume} {97}} (\bibinfo
  {year} {2018})}\BibitemShut {NoStop}%
\bibitem [{\citenamefont {Guo}\ \emph {et~al.}(2019)\citenamefont {Guo},
  \citenamefont {Ye}, \citenamefont {Zhong},\ and\ \citenamefont
  {Liao}}]{guo2019continuous-variable}%
  \BibitemOpen
  \bibfield  {author} {\bibinfo {author} {\bibfnamefont {Y.}~\bibnamefont
  {Guo}}, \bibinfo {author} {\bibfnamefont {W.}~\bibnamefont {Ye}}, \bibinfo
  {author} {\bibfnamefont {H.}~\bibnamefont {Zhong}},\ and\ \bibinfo {author}
  {\bibfnamefont {Q.}~\bibnamefont {Liao}},\ }\bibfield  {title} {\bibinfo
  {title} {Continuous-variable quantum key distribution with non-{G}aussian
  quantum catalysis},\ }\href@noop {} {\bibfield  {journal} {\bibinfo
  {journal} {Physical Review A}\ }\textbf {\bibinfo {volume} {99}} (\bibinfo
  {year} {2019})}\BibitemShut {NoStop}%
\bibitem [{\citenamefont {Hu}\ \emph {et~al.}(2020)\citenamefont {Hu},
  \citenamefont {Al-amri}, \citenamefont {Liao},\ and\ \citenamefont
  {Zubairy}}]{PhysRevA.102.012608}%
  \BibitemOpen
  \bibfield  {author} {\bibinfo {author} {\bibfnamefont {L.}~\bibnamefont
  {Hu}}, \bibinfo {author} {\bibfnamefont {M.}~\bibnamefont {Al-amri}},
  \bibinfo {author} {\bibfnamefont {Z.}~\bibnamefont {Liao}},\ and\ \bibinfo
  {author} {\bibfnamefont {M.~S.}\ \bibnamefont {Zubairy}},\ }\bibfield
  {title} {\bibinfo {title} {Continuous-variable quantum key distribution with
  non-{G}aussian operations},\ }\href@noop {} {\bibfield  {journal} {\bibinfo
  {journal} {Physical Review A}\ }\textbf {\bibinfo {volume} {102}},\ \bibinfo
  {pages} {012608} (\bibinfo {year} {2020})}\BibitemShut {NoStop}%
\bibitem [{\citenamefont {He}\ \emph {et~al.}(2020)\citenamefont {He},
  \citenamefont {Malaney},\ and\ \citenamefont {Green}}]{he2019multi}%
  \BibitemOpen
  \bibfield  {author} {\bibinfo {author} {\bibfnamefont {M.}~\bibnamefont
  {He}}, \bibinfo {author} {\bibfnamefont {R.}~\bibnamefont {Malaney}},\ and\
  \bibinfo {author} {\bibfnamefont {J.}~\bibnamefont {Green}},\ }\bibfield
  {title} {\bibinfo {title} {Multi-mode {CV-QKD} with non-{G}aussian
  operations},\ }\href@noop {} {\bibfield  {journal} {\bibinfo  {journal}
  {Quantum Engineering}\ }\textbf {\bibinfo {volume} {2}},\ \bibinfo {pages}
  {e40} (\bibinfo {year} {2020})}\BibitemShut {NoStop}%
\bibitem [{\citenamefont {Singh}\ and\ \citenamefont
  {Bose}(2021)}]{singh2021non}%
  \BibitemOpen
  \bibfield  {author} {\bibinfo {author} {\bibfnamefont {J.}~\bibnamefont
  {Singh}}\ and\ \bibinfo {author} {\bibfnamefont {S.}~\bibnamefont {Bose}},\
  }\bibfield  {title} {\bibinfo {title} {Non-{G}aussian operations in
  measurement-device-independent quantum key distribution},\ }\href@noop {}
  {\bibfield  {journal} {\bibinfo  {journal} {Physical Review A}\ }\textbf
  {\bibinfo {volume} {104}},\ \bibinfo {pages} {052605} (\bibinfo {year}
  {2021})}\BibitemShut {NoStop}%
\bibitem [{\citenamefont {Chen}\ \emph {et~al.}(2023)\citenamefont {Chen},
  \citenamefont {Jia}, \citenamefont {Zhao}, \citenamefont {Zhou},
  \citenamefont {Liu},\ and\ \citenamefont {Hu}}]{chen2023continuous}%
  \BibitemOpen
  \bibfield  {author} {\bibinfo {author} {\bibfnamefont {X.}~\bibnamefont
  {Chen}}, \bibinfo {author} {\bibfnamefont {F.}~\bibnamefont {Jia}}, \bibinfo
  {author} {\bibfnamefont {T.}~\bibnamefont {Zhao}}, \bibinfo {author}
  {\bibfnamefont {N.}~\bibnamefont {Zhou}}, \bibinfo {author} {\bibfnamefont
  {S.}~\bibnamefont {Liu}},\ and\ \bibinfo {author} {\bibfnamefont
  {L.}~\bibnamefont {Hu}},\ }\bibfield  {title} {\bibinfo {title}
  {Continuous-variable quantum key distribution based on non-{G}aussian
  operations with on-off detection},\ }\href@noop {} {\bibfield  {journal}
  {\bibinfo  {journal} {Optics Express}\ }\textbf {\bibinfo {volume} {31}},\
  \bibinfo {pages} {32935} (\bibinfo {year} {2023})}\BibitemShut {NoStop}%
\bibitem [{\citenamefont {Wang}\ \emph {et~al.}(2023)\citenamefont {Wang},
  \citenamefont {Xu}, \citenamefont {Zhao}, \citenamefont {Chen}, \citenamefont
  {Yu},\ and\ \citenamefont {Guo}}]{wang2023non}%
  \BibitemOpen
  \bibfield  {author} {\bibinfo {author} {\bibfnamefont {X.}~\bibnamefont
  {Wang}}, \bibinfo {author} {\bibfnamefont {M.}~\bibnamefont {Xu}}, \bibinfo
  {author} {\bibfnamefont {Y.}~\bibnamefont {Zhao}}, \bibinfo {author}
  {\bibfnamefont {Z.}~\bibnamefont {Chen}}, \bibinfo {author} {\bibfnamefont
  {S.}~\bibnamefont {Yu}},\ and\ \bibinfo {author} {\bibfnamefont
  {H.}~\bibnamefont {Guo}},\ }\bibfield  {title} {\bibinfo {title}
  {Non-{G}aussian reconciliation for continuous-variable quantum key
  distribution},\ }\href@noop {} {\bibfield  {journal} {\bibinfo  {journal}
  {Physical Review Applied}\ }\textbf {\bibinfo {volume} {19}},\ \bibinfo
  {pages} {054084} (\bibinfo {year} {2023})}\BibitemShut {NoStop}%
\bibitem [{\citenamefont {Ralph}\ and\ \citenamefont
  {Lund}(2009)}]{ralph2009nondeterministic}%
  \BibitemOpen
  \bibfield  {author} {\bibinfo {author} {\bibfnamefont {T.}~\bibnamefont
  {Ralph}}\ and\ \bibinfo {author} {\bibfnamefont {A.}~\bibnamefont {Lund}},\
  }\bibfield  {title} {\bibinfo {title} {Nondeterministic noiseless linear
  amplification of quantum systems},\ }in\ \href@noop {} {\emph {\bibinfo
  {booktitle} {AIP Conference Proceedings}}},\ Vol.\ \bibinfo {volume} {1110}\
  (\bibinfo {organization} {American Institute of Physics},\ \bibinfo {year}
  {2009})\ pp.\ \bibinfo {pages} {155--160}\BibitemShut {NoStop}%
\bibitem [{\citenamefont {Kim}\ \emph {et~al.}(2012)\citenamefont {Kim},
  \citenamefont {Lee}, \citenamefont {Ji},\ and\ \citenamefont
  {Nha}}]{kim2012quantum}%
  \BibitemOpen
  \bibfield  {author} {\bibinfo {author} {\bibfnamefont {H.}~\bibnamefont
  {Kim}}, \bibinfo {author} {\bibfnamefont {S.}~\bibnamefont {Lee}}, \bibinfo
  {author} {\bibfnamefont {S.}~\bibnamefont {Ji}},\ and\ \bibinfo {author}
  {\bibfnamefont {H.}~\bibnamefont {Nha}},\ }\bibfield  {title} {\bibinfo
  {title} {Quantum linear amplifier enhanced by photon subtraction and
  addition},\ }\href@noop {} {\bibfield  {journal} {\bibinfo  {journal}
  {Physical Review A}\ }\textbf {\bibinfo {volume} {85}},\ \bibinfo {pages}
  {013839} (\bibinfo {year} {2012})}\BibitemShut {NoStop}%
\bibitem [{\citenamefont {Yang}\ \emph {et~al.}(2012)\citenamefont {Yang},
  \citenamefont {Zhang}, \citenamefont {Zou}, \citenamefont {Bi},\ and\
  \citenamefont {Lin}}]{yang2012continuous-variable}%
  \BibitemOpen
  \bibfield  {author} {\bibinfo {author} {\bibfnamefont {S.}~\bibnamefont
  {Yang}}, \bibinfo {author} {\bibfnamefont {S.}~\bibnamefont {Zhang}},
  \bibinfo {author} {\bibfnamefont {X.}~\bibnamefont {Zou}}, \bibinfo {author}
  {\bibfnamefont {S.}~\bibnamefont {Bi}},\ and\ \bibinfo {author}
  {\bibfnamefont {X.}~\bibnamefont {Lin}},\ }\bibfield  {title} {\bibinfo
  {title} {Continuous-variable entanglement distillation with noiseless linear
  amplification},\ }\href@noop {} {\bibfield  {journal} {\bibinfo  {journal}
  {Physical Review A}\ }\textbf {\bibinfo {volume} {86}},\ \bibinfo {pages}
  {062321} (\bibinfo {year} {2012})}\BibitemShut {NoStop}%
\bibitem [{\citenamefont {Gagatsos}\ \emph {et~al.}(2014)\citenamefont
  {Gagatsos}, \citenamefont {Fiurášek}, \citenamefont {Zavatta},
  \citenamefont {Bellini},\ and\ \citenamefont {Cerf}}]{gagatsos2014heralded}%
  \BibitemOpen
  \bibfield  {author} {\bibinfo {author} {\bibfnamefont {C.~N.}\ \bibnamefont
  {Gagatsos}}, \bibinfo {author} {\bibfnamefont {J.}~\bibnamefont
  {Fiurášek}}, \bibinfo {author} {\bibfnamefont {A.}~\bibnamefont {Zavatta}},
  \bibinfo {author} {\bibfnamefont {M.}~\bibnamefont {Bellini}},\ and\ \bibinfo
  {author} {\bibfnamefont {N.~J.}\ \bibnamefont {Cerf}},\ }\bibfield  {title}
  {\bibinfo {title} {Heralded noiseless amplification and attenuation of
  non-{G}aussian states of light},\ }\href@noop {} {\bibfield  {journal}
  {\bibinfo  {journal} {Physical Review A}\ }\textbf {\bibinfo {volume} {89}}
  (\bibinfo {year} {2014})}\BibitemShut {NoStop}%
\bibitem [{\citenamefont {Zhang}\ and\ \citenamefont
  {Zhang}(2018)}]{zhang2018photon}%
  \BibitemOpen
  \bibfield  {author} {\bibinfo {author} {\bibfnamefont {S.}~\bibnamefont
  {Zhang}}\ and\ \bibinfo {author} {\bibfnamefont {X.}~\bibnamefont {Zhang}},\
  }\bibfield  {title} {\bibinfo {title} {Photon catalysis acting as noiseless
  linear amplification and its application in coherence enhancement},\
  }\href@noop {} {\bibfield  {journal} {\bibinfo  {journal} {Physical Review
  A}\ }\textbf {\bibinfo {volume} {97}} (\bibinfo {year} {2018})}\BibitemShut
  {NoStop}%
\bibitem [{\citenamefont {Adnane}\ \emph {et~al.}(2019)\citenamefont {Adnane},
  \citenamefont {Bina}, \citenamefont {Albarelli}, \citenamefont {Gharbi},\
  and\ \citenamefont {Paris}}]{adnane2019quantum}%
  \BibitemOpen
  \bibfield  {author} {\bibinfo {author} {\bibfnamefont {H.}~\bibnamefont
  {Adnane}}, \bibinfo {author} {\bibfnamefont {M.}~\bibnamefont {Bina}},
  \bibinfo {author} {\bibfnamefont {F.}~\bibnamefont {Albarelli}}, \bibinfo
  {author} {\bibfnamefont {A.}~\bibnamefont {Gharbi}},\ and\ \bibinfo {author}
  {\bibfnamefont {M.~G.~A.}\ \bibnamefont {Paris}},\ }\bibfield  {title}
  {\bibinfo {title} {Quantum state engineering by nondeterministic noiseless
  linear amplification},\ }\href@noop {} {\bibfield  {journal} {\bibinfo
  {journal} {Physical Review A}\ }\textbf {\bibinfo {volume} {99}} (\bibinfo
  {year} {2019})}\BibitemShut {NoStop}%
\bibitem [{\citenamefont {Fiurášek}(2022)}]{2022_443}%
  \BibitemOpen
  \bibfield  {author} {\bibinfo {author} {\bibfnamefont {J.}~\bibnamefont
  {Fiurášek}},\ }\bibfield  {title} {\bibinfo {title} {Teleportation-based
  noiseless quantum amplification of coherent states of light},\ }\bibfield
  {journal} {\bibinfo  {journal} {Optics Express}\ }\textbf {\bibinfo {volume}
  {30}},\ \href {https://doi.org/10.1364/oe.443389} {10.1364/oe.443389}
  (\bibinfo {year} {2022})\BibitemShut {NoStop}%
\bibitem [{\citenamefont {Lund}\ and\ \citenamefont
  {Ralph}(2009)}]{lund2009continuous-variable}%
  \BibitemOpen
  \bibfield  {author} {\bibinfo {author} {\bibfnamefont {A.~P.}\ \bibnamefont
  {Lund}}\ and\ \bibinfo {author} {\bibfnamefont {T.~C.}\ \bibnamefont
  {Ralph}},\ }\bibfield  {title} {\bibinfo {title} {Continuous-variable
  entanglement distillation over a general lossy channel},\ }\href@noop {}
  {\bibfield  {journal} {\bibinfo  {journal} {Physical Review A}\ }\textbf
  {\bibinfo {volume} {80}},\ \bibinfo {pages} {032309} (\bibinfo {year}
  {2009})}\BibitemShut {NoStop}%
\bibitem [{\citenamefont {Zhang}\ and\ \citenamefont
  {Van~Loock}(2010)}]{zhang2010distillation}%
  \BibitemOpen
  \bibfield  {author} {\bibinfo {author} {\bibfnamefont {S.~L.}\ \bibnamefont
  {Zhang}}\ and\ \bibinfo {author} {\bibfnamefont {P.}~\bibnamefont
  {Van~Loock}},\ }\bibfield  {title} {\bibinfo {title} {Distillation of
  mixed-state continuous-variable entanglement by photon subtraction},\
  }\href@noop {} {\bibfield  {journal} {\bibinfo  {journal} {Physical Review
  A}\ }\textbf {\bibinfo {volume} {82}},\ \bibinfo {pages} {062316} (\bibinfo
  {year} {2010})}\BibitemShut {NoStop}%
\bibitem [{\citenamefont {Fiurášek}(2010)}]{fiurasek2010distillation}%
  \BibitemOpen
  \bibfield  {author} {\bibinfo {author} {\bibfnamefont {J.}~\bibnamefont
  {Fiurášek}},\ }\bibfield  {title} {\bibinfo {title} {Distillation and
  purification of symmetric entangled {G}aussian states},\ }\href@noop {}
  {\bibfield  {journal} {\bibinfo  {journal} {Physical Review A}\ }\textbf
  {\bibinfo {volume} {82}} (\bibinfo {year} {2010})}\BibitemShut {NoStop}%
\bibitem [{\citenamefont {Datta}\ \emph {et~al.}(2012)\citenamefont {Datta},
  \citenamefont {Zhang}, \citenamefont {Nunn}, \citenamefont {Langford},
  \citenamefont {Feito}, \citenamefont {Plenio},\ and\ \citenamefont
  {Walmsley}}]{datta2012compact}%
  \BibitemOpen
  \bibfield  {author} {\bibinfo {author} {\bibfnamefont {A.}~\bibnamefont
  {Datta}}, \bibinfo {author} {\bibfnamefont {L.}~\bibnamefont {Zhang}},
  \bibinfo {author} {\bibfnamefont {J.}~\bibnamefont {Nunn}}, \bibinfo {author}
  {\bibfnamefont {N.~K.}\ \bibnamefont {Langford}}, \bibinfo {author}
  {\bibfnamefont {A.}~\bibnamefont {Feito}}, \bibinfo {author} {\bibfnamefont
  {M.~B.}\ \bibnamefont {Plenio}},\ and\ \bibinfo {author} {\bibfnamefont
  {I.~A.}\ \bibnamefont {Walmsley}},\ }\bibfield  {title} {\bibinfo {title}
  {Compact continuous-variable entanglement distillation.},\ }\href@noop {}
  {\bibfield  {journal} {\bibinfo  {journal} {Physical Review Letters}\
  }\textbf {\bibinfo {volume} {108}},\ \bibinfo {pages} {060502} (\bibinfo
  {year} {2012})}\BibitemShut {NoStop}%
\bibitem [{\citenamefont {Lee}\ and\ \citenamefont
  {Nha}(2013)}]{lee2013entanglement}%
  \BibitemOpen
  \bibfield  {author} {\bibinfo {author} {\bibfnamefont {J.}~\bibnamefont
  {Lee}}\ and\ \bibinfo {author} {\bibfnamefont {H.}~\bibnamefont {Nha}},\
  }\bibfield  {title} {\bibinfo {title} {Entanglement distillation for
  continuous variables in a thermal environment: Effectiveness of a
  non-{G}aussian operation},\ }\href@noop {} {\bibfield  {journal} {\bibinfo
  {journal} {Physical Review A}\ }\textbf {\bibinfo {volume} {87}},\ \bibinfo
  {pages} {032307} (\bibinfo {year} {2013})}\BibitemShut {NoStop}%
\bibitem [{\citenamefont {Zhang}\ \emph {et~al.}(2013)\citenamefont {Zhang},
  \citenamefont {Guo}, \citenamefont {Zou}, \citenamefont {Shi},\ and\
  \citenamefont {Dong}}]{zhang2013continuous-variable-entanglement}%
  \BibitemOpen
  \bibfield  {author} {\bibinfo {author} {\bibfnamefont {S.~L.}\ \bibnamefont
  {Zhang}}, \bibinfo {author} {\bibfnamefont {G.}~\bibnamefont {Guo}}, \bibinfo
  {author} {\bibfnamefont {X.}~\bibnamefont {Zou}}, \bibinfo {author}
  {\bibfnamefont {B.}~\bibnamefont {Shi}},\ and\ \bibinfo {author}
  {\bibfnamefont {Y.}~\bibnamefont {Dong}},\ }\bibfield  {title} {\bibinfo
  {title} {Continuous-variable-entanglement distillation with photon
  addition},\ }\href@noop {} {\bibfield  {journal} {\bibinfo  {journal}
  {Physical Review A}\ }\textbf {\bibinfo {volume} {88}} (\bibinfo {year}
  {2013})}\BibitemShut {NoStop}%
\bibitem [{\citenamefont {Seshadreesan}\ \emph {et~al.}(2019)\citenamefont
  {Seshadreesan}, \citenamefont {Krovi},\ and\ \citenamefont
  {Guha}}]{seshadreesan2019continuous-variable}%
  \BibitemOpen
  \bibfield  {author} {\bibinfo {author} {\bibfnamefont {K.~P.}\ \bibnamefont
  {Seshadreesan}}, \bibinfo {author} {\bibfnamefont {H.}~\bibnamefont
  {Krovi}},\ and\ \bibinfo {author} {\bibfnamefont {S.}~\bibnamefont {Guha}},\
  }\bibfield  {title} {\bibinfo {title} {Continuous-variable entanglement
  distillation over a pure loss channel with multiple quantum scissors},\
  }\href@noop {} {\bibfield  {journal} {\bibinfo  {journal} {Physical Review
  A}\ }\textbf {\bibinfo {volume} {100}} (\bibinfo {year} {2019})}\BibitemShut
  {NoStop}%
\bibitem [{\citenamefont {Mardani}\ \emph {et~al.}(2020)\citenamefont
  {Mardani}, \citenamefont {Shafiei}, \citenamefont {Ghadimi},\ and\
  \citenamefont {Abdi}}]{mardani2020continuous}%
  \BibitemOpen
  \bibfield  {author} {\bibinfo {author} {\bibfnamefont {Y.}~\bibnamefont
  {Mardani}}, \bibinfo {author} {\bibfnamefont {A.}~\bibnamefont {Shafiei}},
  \bibinfo {author} {\bibfnamefont {M.}~\bibnamefont {Ghadimi}},\ and\ \bibinfo
  {author} {\bibfnamefont {M.}~\bibnamefont {Abdi}},\ }\bibfield  {title}
  {\bibinfo {title} {Continuous-variable entanglement distillation by cascaded
  photon replacement},\ }\href@noop {} {\bibfield  {journal} {\bibinfo
  {journal} {Physical Review A}\ }\textbf {\bibinfo {volume} {102}},\ \bibinfo
  {pages} {012407} (\bibinfo {year} {2020})}\BibitemShut {NoStop}%
\bibitem [{\citenamefont {Ghose}\ and\ \citenamefont
  {Sanders}(2007)}]{ghose2007non}%
  \BibitemOpen
  \bibfield  {author} {\bibinfo {author} {\bibfnamefont {S.}~\bibnamefont
  {Ghose}}\ and\ \bibinfo {author} {\bibfnamefont {B.~C.}\ \bibnamefont
  {Sanders}},\ }\bibfield  {title} {\bibinfo {title} {Non-{G}aussian ancilla
  states for continuous variable quantum computation via {G}aussian maps},\
  }\href@noop {} {\bibfield  {journal} {\bibinfo  {journal} {Journal of Modern
  Optics}\ }\textbf {\bibinfo {volume} {54}},\ \bibinfo {pages} {855} (\bibinfo
  {year} {2007})}\BibitemShut {NoStop}%
\bibitem [{\citenamefont {Marshall}\ \emph {et~al.}(2015)\citenamefont
  {Marshall}, \citenamefont {Pooser}, \citenamefont {Siopsis},\ and\
  \citenamefont {Weedbrook}}]{marshall2015repeat}%
  \BibitemOpen
  \bibfield  {author} {\bibinfo {author} {\bibfnamefont {K.}~\bibnamefont
  {Marshall}}, \bibinfo {author} {\bibfnamefont {R.}~\bibnamefont {Pooser}},
  \bibinfo {author} {\bibfnamefont {G.}~\bibnamefont {Siopsis}},\ and\ \bibinfo
  {author} {\bibfnamefont {C.}~\bibnamefont {Weedbrook}},\ }\bibfield  {title}
  {\bibinfo {title} {Repeat-until-success cubic phase gate for universal
  continuous-variable quantum computation},\ }\href@noop {} {\bibfield
  {journal} {\bibinfo  {journal} {Physical Review A}\ }\textbf {\bibinfo
  {volume} {91}},\ \bibinfo {pages} {032321} (\bibinfo {year}
  {2015})}\BibitemShut {NoStop}%
\bibitem [{\citenamefont {Miyata}\ \emph {et~al.}(2016)\citenamefont {Miyata},
  \citenamefont {Ogawa}, \citenamefont {Marek}, \citenamefont {Filip},
  \citenamefont {Yonezawa}, \citenamefont {Yoshikawa},\ and\ \citenamefont
  {Furusawa}}]{miyata2016implementation}%
  \BibitemOpen
  \bibfield  {author} {\bibinfo {author} {\bibfnamefont {K.}~\bibnamefont
  {Miyata}}, \bibinfo {author} {\bibfnamefont {H.}~\bibnamefont {Ogawa}},
  \bibinfo {author} {\bibfnamefont {P.}~\bibnamefont {Marek}}, \bibinfo
  {author} {\bibfnamefont {R.}~\bibnamefont {Filip}}, \bibinfo {author}
  {\bibfnamefont {H.}~\bibnamefont {Yonezawa}}, \bibinfo {author}
  {\bibfnamefont {J.-i.}\ \bibnamefont {Yoshikawa}},\ and\ \bibinfo {author}
  {\bibfnamefont {A.}~\bibnamefont {Furusawa}},\ }\bibfield  {title} {\bibinfo
  {title} {Implementation of a quantum cubic gate by an adaptive non-{G}aussian
  measurement},\ }\href@noop {} {\bibfield  {journal} {\bibinfo  {journal}
  {Physical Review A}\ }\textbf {\bibinfo {volume} {93}},\ \bibinfo {pages}
  {022301} (\bibinfo {year} {2016})}\BibitemShut {NoStop}%
\bibitem [{\citenamefont {Marshall}\ \emph {et~al.}(2016)\citenamefont
  {Marshall}, \citenamefont {Jacobsen}, \citenamefont {Sch{\"a}fermeier},
  \citenamefont {Gehring}, \citenamefont {Weedbrook},\ and\ \citenamefont
  {Andersen}}]{marshall2016continuous}%
  \BibitemOpen
  \bibfield  {author} {\bibinfo {author} {\bibfnamefont {K.}~\bibnamefont
  {Marshall}}, \bibinfo {author} {\bibfnamefont {C.~S.}\ \bibnamefont
  {Jacobsen}}, \bibinfo {author} {\bibfnamefont {C.}~\bibnamefont
  {Sch{\"a}fermeier}}, \bibinfo {author} {\bibfnamefont {T.}~\bibnamefont
  {Gehring}}, \bibinfo {author} {\bibfnamefont {C.}~\bibnamefont {Weedbrook}},\
  and\ \bibinfo {author} {\bibfnamefont {U.~L.}\ \bibnamefont {Andersen}},\
  }\bibfield  {title} {\bibinfo {title} {Continuous-variable quantum computing
  on encrypted data},\ }\href@noop {} {\bibfield  {journal} {\bibinfo
  {journal} {Nature communications}\ }\textbf {\bibinfo {volume} {7}},\
  \bibinfo {pages} {1} (\bibinfo {year} {2016})}\BibitemShut {NoStop}%
\bibitem [{\citenamefont {Cochrane}\ \emph {et~al.}(2002)\citenamefont
  {Cochrane}, \citenamefont {Ralph},\ and\ \citenamefont {Milburn}}]{2002_188}%
  \BibitemOpen
  \bibfield  {author} {\bibinfo {author} {\bibfnamefont {P.~T.}\ \bibnamefont
  {Cochrane}}, \bibinfo {author} {\bibfnamefont {T.~C.}\ \bibnamefont
  {Ralph}},\ and\ \bibinfo {author} {\bibfnamefont {G.~J.}\ \bibnamefont
  {Milburn}},\ }\bibfield  {title} {\bibinfo {title} {Teleportation improvement
  by conditional measurements on the two-mode squeezed vacuum},\ }\bibfield
  {journal} {\bibinfo  {journal} {Physical Review A}\ }\textbf {\bibinfo
  {volume} {65}},\ \href {https://doi.org/10.1103/PhysRevA.65.062306}
  {10.1103/PhysRevA.65.062306} (\bibinfo {year} {2002})\BibitemShut {NoStop}%
\bibitem [{\citenamefont {Dell’Anno}\ \emph {et~al.}(2007)\citenamefont
  {Dell’Anno}, \citenamefont {De~Siena}, \citenamefont {Albano},\ and\
  \citenamefont {Illuminati}}]{2007_189}%
  \BibitemOpen
  \bibfield  {author} {\bibinfo {author} {\bibfnamefont {F.}~\bibnamefont
  {Dell’Anno}}, \bibinfo {author} {\bibfnamefont {S.}~\bibnamefont
  {De~Siena}}, \bibinfo {author} {\bibfnamefont {L.}~\bibnamefont {Albano}},\
  and\ \bibinfo {author} {\bibfnamefont {F.}~\bibnamefont {Illuminati}},\
  }\bibfield  {title} {\bibinfo {title} {Continuous-variable quantum
  teleportation with non-{G}aussian resources},\ }\bibfield  {journal}
  {\bibinfo  {journal} {Physical Review A}\ }\textbf {\bibinfo {volume} {76}},\
  \href {https://doi.org/10.1103/PhysRevA.76.022301}
  {10.1103/PhysRevA.76.022301} (\bibinfo {year} {2007})\BibitemShut {NoStop}%
\bibitem [{\citenamefont {Yang}\ and\ \citenamefont {Li}(2009)}]{2009_183}%
  \BibitemOpen
  \bibfield  {author} {\bibinfo {author} {\bibfnamefont {Y.}~\bibnamefont
  {Yang}}\ and\ \bibinfo {author} {\bibfnamefont {F.-L.}\ \bibnamefont {Li}},\
  }\bibfield  {title} {\bibinfo {title} {Entanglement properties of
  non-{G}aussian resources generated via photon subtraction and addition and
  continuous-variable quantum-teleportation improvement},\ }\bibfield
  {journal} {\bibinfo  {journal} {Physical Review A}\ }\textbf {\bibinfo
  {volume} {80}},\ \href {https://doi.org/10.1103/PhysRevA.80.022315}
  {10.1103/PhysRevA.80.022315} (\bibinfo {year} {2009})\BibitemShut {NoStop}%
\bibitem [{\citenamefont {Dell’Anno}\ \emph {et~al.}(2010)\citenamefont
  {Dell’Anno}, \citenamefont {De~Siena},\ and\ \citenamefont
  {Illuminati}}]{2010_333}%
  \BibitemOpen
  \bibfield  {author} {\bibinfo {author} {\bibfnamefont {F.}~\bibnamefont
  {Dell’Anno}}, \bibinfo {author} {\bibfnamefont {S.}~\bibnamefont
  {De~Siena}},\ and\ \bibinfo {author} {\bibfnamefont {F.}~\bibnamefont
  {Illuminati}},\ }\bibfield  {title} {\bibinfo {title} {Realistic
  continuous-variable quantum teleportation with non-{G}aussian resources},\
  }\bibfield  {journal} {\bibinfo  {journal} {Physical Review A}\ }\textbf
  {\bibinfo {volume} {81}},\ \href {https://doi.org/10.1103/PhysRevA.81.012333}
  {10.1103/PhysRevA.81.012333} (\bibinfo {year} {2010})\BibitemShut {NoStop}%
\bibitem [{\citenamefont {Seshadreesan}\ \emph {et~al.}(2015)\citenamefont
  {Seshadreesan}, \citenamefont {Dowling},\ and\ \citenamefont
  {Agarwal}}]{2015_297}%
  \BibitemOpen
  \bibfield  {author} {\bibinfo {author} {\bibfnamefont {K.~P.}\ \bibnamefont
  {Seshadreesan}}, \bibinfo {author} {\bibfnamefont {J.~P.}\ \bibnamefont
  {Dowling}},\ and\ \bibinfo {author} {\bibfnamefont {G.~S.}\ \bibnamefont
  {Agarwal}},\ }\bibfield  {title} {\bibinfo {title} {Non-{G}aussian entangled
  states and quantum teleportation of schrödinger-cat states},\ }\bibfield
  {journal} {\bibinfo  {journal} {Physica Scripta}\ }\textbf {\bibinfo {volume}
  {90}},\ \href {https://doi.org/10.1088/0031-8949/90/7/074029}
  {10.1088/0031-8949/90/7/074029} (\bibinfo {year} {2015})\BibitemShut
  {NoStop}%
\bibitem [{\citenamefont {Xu}(2015)}]{2015_325}%
  \BibitemOpen
  \bibfield  {author} {\bibinfo {author} {\bibfnamefont {X.-x.}\ \bibnamefont
  {Xu}},\ }\bibfield  {title} {\bibinfo {title} {Enhancing quantum entanglement
  and quantum teleportation for two-mode squeezed vacuum state by local
  quantum-optical catalysis},\ }\bibfield  {journal} {\bibinfo  {journal}
  {Physical Review A}\ }\textbf {\bibinfo {volume} {92}},\ \href
  {https://doi.org/10.1103/PhysRevA.92.012318} {10.1103/PhysRevA.92.012318}
  (\bibinfo {year} {2015})\BibitemShut {NoStop}%
\bibitem [{\citenamefont {Wang}\ \emph {et~al.}(2015)\citenamefont {Wang},
  \citenamefont {Hou}, \citenamefont {Chen},\ and\ \citenamefont
  {Xu}}]{2015_334}%
  \BibitemOpen
  \bibfield  {author} {\bibinfo {author} {\bibfnamefont {S.}~\bibnamefont
  {Wang}}, \bibinfo {author} {\bibfnamefont {L.-L.}\ \bibnamefont {Hou}},
  \bibinfo {author} {\bibfnamefont {X.-F.}\ \bibnamefont {Chen}},\ and\
  \bibinfo {author} {\bibfnamefont {X.-F.}\ \bibnamefont {Xu}},\ }\bibfield
  {title} {\bibinfo {title} {Continuous-variable quantum teleportation with
  non-{G}aussian entangled states generated via multiple-photon subtraction and
  addition},\ }\bibfield  {journal} {\bibinfo  {journal} {Physical Review A}\
  }\textbf {\bibinfo {volume} {91}},\ \href
  {https://doi.org/10.1103/PhysRevA.91.063832} {10.1103/PhysRevA.91.063832}
  (\bibinfo {year} {2015})\BibitemShut {NoStop}%
\bibitem [{\citenamefont {Hu}\ \emph {et~al.}(2017)\citenamefont {Hu},
  \citenamefont {Liao},\ and\ \citenamefont {Zubairy}}]{2017_178}%
  \BibitemOpen
  \bibfield  {author} {\bibinfo {author} {\bibfnamefont {L.}~\bibnamefont
  {Hu}}, \bibinfo {author} {\bibfnamefont {Z.}~\bibnamefont {Liao}},\ and\
  \bibinfo {author} {\bibfnamefont {M.~S.}\ \bibnamefont {Zubairy}},\
  }\bibfield  {title} {\bibinfo {title} {Continuous-variable entanglement via
  multiphoton catalysis},\ }\bibfield  {journal} {\bibinfo  {journal} {Physical
  Review A}\ }\textbf {\bibinfo {volume} {95}},\ \href
  {https://doi.org/10.1103/PhysRevA.95.012310} {10.1103/PhysRevA.95.012310}
  (\bibinfo {year} {2017})\BibitemShut {NoStop}%
\bibitem [{\citenamefont {Ye}\ \emph {et~al.}(2020)\citenamefont {Ye},
  \citenamefont {Guo}, \citenamefont {Zhang}, \citenamefont {Zhong},
  \citenamefont {Xia}, \citenamefont {Chang},\ and\ \citenamefont
  {Hu}}]{ye2020nonclassicality}%
  \BibitemOpen
  \bibfield  {author} {\bibinfo {author} {\bibfnamefont {W.}~\bibnamefont
  {Ye}}, \bibinfo {author} {\bibfnamefont {Y.}~\bibnamefont {Guo}}, \bibinfo
  {author} {\bibfnamefont {H.}~\bibnamefont {Zhang}}, \bibinfo {author}
  {\bibfnamefont {H.}~\bibnamefont {Zhong}}, \bibinfo {author} {\bibfnamefont
  {Y.}~\bibnamefont {Xia}}, \bibinfo {author} {\bibfnamefont {S.}~\bibnamefont
  {Chang}},\ and\ \bibinfo {author} {\bibfnamefont {L.}~\bibnamefont {Hu}},\
  }\bibfield  {title} {\bibinfo {title} {Nonclassicality and entanglement
  properties of non-{G}aussian entangled states via a superposition of
  number-conserving operations},\ }\href@noop {} {\bibfield  {journal}
  {\bibinfo  {journal} {Quantum Information Processing}\ }\textbf {\bibinfo
  {volume} {19}},\ \bibinfo {pages} {1} (\bibinfo {year} {2020})}\BibitemShut
  {NoStop}%
\bibitem [{\citenamefont {Meng}\ \emph {et~al.}(2020)\citenamefont {Meng},
  \citenamefont {Li}, \citenamefont {Wang}, \citenamefont {Zhang},
  \citenamefont {Zhang}, \citenamefont {Yang},\ and\ \citenamefont
  {Liang}}]{2020_485}%
  \BibitemOpen
  \bibfield  {author} {\bibinfo {author} {\bibfnamefont {X.}~\bibnamefont
  {Meng}}, \bibinfo {author} {\bibfnamefont {K.}~\bibnamefont {Li}}, \bibinfo
  {author} {\bibfnamefont {J.}~\bibnamefont {Wang}}, \bibinfo {author}
  {\bibfnamefont {X.}~\bibnamefont {Zhang}}, \bibinfo {author} {\bibfnamefont
  {Z.}~\bibnamefont {Zhang}}, \bibinfo {author} {\bibfnamefont
  {Z.}~\bibnamefont {Yang}},\ and\ \bibinfo {author} {\bibfnamefont
  {B.}~\bibnamefont {Liang}},\ }\bibfield  {title} {\bibinfo {title}
  {Continuous‐variable entanglement and wigner‐function negativity via
  adding or subtracting photons},\ }\bibfield  {journal} {\bibinfo  {journal}
  {Annalen der Physik}\ }\textbf {\bibinfo {volume} {532}},\ \href
  {https://doi.org/10.1002/andp.201900585} {10.1002/andp.201900585} (\bibinfo
  {year} {2020})\BibitemShut {NoStop}%
\bibitem [{\citenamefont {Bose}\ and\ \citenamefont {Kumar}(2021)}]{2021_380}%
  \BibitemOpen
  \bibfield  {author} {\bibinfo {author} {\bibfnamefont {S.}~\bibnamefont
  {Bose}}\ and\ \bibinfo {author} {\bibfnamefont {M.~S.}\ \bibnamefont
  {Kumar}},\ }\bibfield  {title} {\bibinfo {title} {Analysis of necessary and
  sufficient conditions for quantum teleportation with non-{G}aussian
  resources},\ }\bibfield  {journal} {\bibinfo  {journal} {Physical Review A}\
  }\textbf {\bibinfo {volume} {103}},\ \href
  {https://doi.org/10.1103/PhysRevA.103.032432} {10.1103/PhysRevA.103.032432}
  (\bibinfo {year} {2021})\BibitemShut {NoStop}%
\bibitem [{\citenamefont {Villase{\~n}or}\ and\ \citenamefont
  {Malaney}(2021)}]{villasenor2021enhancing}%
  \BibitemOpen
  \bibfield  {author} {\bibinfo {author} {\bibfnamefont {E.}~\bibnamefont
  {Villase{\~n}or}}\ and\ \bibinfo {author} {\bibfnamefont {R.}~\bibnamefont
  {Malaney}},\ }\bibfield  {title} {\bibinfo {title} {Enhancing continuous
  variable quantum teleportation using non-{G}aussian resources},\ }in\
  \href@noop {} {\emph {\bibinfo {booktitle} {2021 IEEE Global Communications
  Conference (GLOBECOM)}}}\ (\bibinfo {organization} {IEEE},\ \bibinfo {year}
  {2021})\ pp.\ \bibinfo {pages} {1--6}\BibitemShut {NoStop}%
\bibitem [{\citenamefont {Dat}\ and\ \citenamefont {Duc}(2022)}]{2022_483}%
  \BibitemOpen
  \bibfield  {author} {\bibinfo {author} {\bibfnamefont {T.~Q.}\ \bibnamefont
  {Dat}}\ and\ \bibinfo {author} {\bibfnamefont {T.~M.}\ \bibnamefont {Duc}},\
  }\bibfield  {title} {\bibinfo {title} {Entanglement, nonlocal features,
  quantum teleportation of two-mode squeezed vacuum states with superposition
  of photon-pair addition and subtraction operations},\ }\bibfield  {journal}
  {\bibinfo  {journal} {Optik}\ }\textbf {\bibinfo {volume} {257}},\ \href
  {https://doi.org/10.1016/j.ijleo.2022.168744} {10.1016/j.ijleo.2022.168744}
  (\bibinfo {year} {2022})\BibitemShut {NoStop}%
\bibitem [{\citenamefont {Kumar}\ and\ \citenamefont {Arora}(2023)}]{2023_475}%
  \BibitemOpen
  \bibfield  {author} {\bibinfo {author} {\bibfnamefont {C.}~\bibnamefont
  {Kumar}}\ and\ \bibinfo {author} {\bibfnamefont {S.}~\bibnamefont {Arora}},\
  }\bibfield  {title} {\bibinfo {title} {Success probability and performance
  optimization in non-{G}aussian continuous-variable quantum teleportation},\
  }\bibfield  {journal} {\bibinfo  {journal} {Physical Review A}\ }\textbf
  {\bibinfo {volume} {107}},\ \href
  {https://doi.org/10.1103/PhysRevA.107.012418} {10.1103/PhysRevA.107.012418}
  (\bibinfo {year} {2023})\BibitemShut {NoStop}%
\bibitem [{\citenamefont {Duc}\ and\ \citenamefont {Dat}(2023)}]{2023_478}%
  \BibitemOpen
  \bibfield  {author} {\bibinfo {author} {\bibfnamefont {T.~M.}\ \bibnamefont
  {Duc}}\ and\ \bibinfo {author} {\bibfnamefont {T.~Q.}\ \bibnamefont {Dat}},\
  }\bibfield  {title} {\bibinfo {title} {Enhanced entanglement and quantum
  teleportation of two-mode squeezed vacuum state via multistage non-{G}aussian
  operations},\ }\bibfield  {journal} {\bibinfo  {journal} {Optik}\ }\textbf
  {\bibinfo {volume} {287}},\ \href
  {https://doi.org/10.1016/j.ijleo.2023.170988} {10.1016/j.ijleo.2023.170988}
  (\bibinfo {year} {2023})\BibitemShut {NoStop}%
\bibitem [{\citenamefont {Chuong}\ and\ \citenamefont {Duc}(2023)}]{2023_486}%
  \BibitemOpen
  \bibfield  {author} {\bibinfo {author} {\bibfnamefont {H.~S.}\ \bibnamefont
  {Chuong}}\ and\ \bibinfo {author} {\bibfnamefont {T.~M.}\ \bibnamefont
  {Duc}},\ }\bibfield  {title} {\bibinfo {title} {Enhancement of
  non-{G}aussianity and nonclassicality of pair coherent states by
  superposition of photon addition and subtraction},\ }\bibfield  {journal}
  {\bibinfo  {journal} {Journal of Physics B: Atomic, Molecular and Optical
  Physics}\ }\textbf {\bibinfo {volume} {56}},\ \href
  {https://doi.org/10.1088/1361-6455/acf484} {10.1088/1361-6455/acf484}
  (\bibinfo {year} {2023})\BibitemShut {NoStop}%
\bibitem [{\citenamefont {Zinatullin}\ \emph {et~al.}(2023)\citenamefont
  {Zinatullin}, \citenamefont {Korolev},\ and\ \citenamefont
  {Golubeva}}]{2023_489}%
  \BibitemOpen
  \bibfield  {author} {\bibinfo {author} {\bibfnamefont {E.~R.}\ \bibnamefont
  {Zinatullin}}, \bibinfo {author} {\bibfnamefont {S.~B.}\ \bibnamefont
  {Korolev}},\ and\ \bibinfo {author} {\bibfnamefont {T.~Y.}\ \bibnamefont
  {Golubeva}},\ }\bibfield  {title} {\bibinfo {title} {Teleportation protocols
  with non-{G}aussian operations: Conditional photon subtraction versus cubic
  phase gate},\ }\bibfield  {journal} {\bibinfo  {journal} {Physical Review A}\
  }\textbf {\bibinfo {volume} {107}},\ \href
  {https://doi.org/10.1103/PhysRevA.107.022422} {10.1103/PhysRevA.107.022422}
  (\bibinfo {year} {2023})\BibitemShut {NoStop}%
\bibitem [{\citenamefont {Sharma}\ \emph {et~al.}(2022)\citenamefont {Sharma},
  \citenamefont {Sanders},\ and\ \citenamefont {Wilde}}]{2022_496}%
  \BibitemOpen
  \bibfield  {author} {\bibinfo {author} {\bibfnamefont {K.}~\bibnamefont
  {Sharma}}, \bibinfo {author} {\bibfnamefont {B.~C.}\ \bibnamefont
  {Sanders}},\ and\ \bibinfo {author} {\bibfnamefont {M.~M.}\ \bibnamefont
  {Wilde}},\ }\bibfield  {title} {\bibinfo {title} {Optimal tests for
  continuous-variable quantum teleportation and photodetectors},\ }\bibfield
  {journal} {\bibinfo  {journal} {Physical Review Research}\ }\textbf {\bibinfo
  {volume} {4}},\ \href {https://doi.org/10.1103/PhysRevResearch.4.023066}
  {10.1103/PhysRevResearch.4.023066} (\bibinfo {year} {2022})\BibitemShut
  {NoStop}%
\bibitem [{\citenamefont {Mishra}\ \emph {et~al.}(2023)\citenamefont {Mishra},
  \citenamefont {Oskouei},\ and\ \citenamefont {Wilde}}]{2023_494}%
  \BibitemOpen
  \bibfield  {author} {\bibinfo {author} {\bibfnamefont {H.~K.}\ \bibnamefont
  {Mishra}}, \bibinfo {author} {\bibfnamefont {S.~K.}\ \bibnamefont
  {Oskouei}},\ and\ \bibinfo {author} {\bibfnamefont {M.~M.}\ \bibnamefont
  {Wilde}},\ }\bibfield  {title} {\bibinfo {title} {Optimal input states for
  quantifying the performance of {CV} unidirectional and bidirectional
  teleportation},\ }\bibfield  {journal} {\bibinfo  {journal} {Physical Review
  A}\ }\textbf {\bibinfo {volume} {107}},\ \href
  {https://doi.org/10.1103/PhysRevA.107.062603} {10.1103/PhysRevA.107.062603}
  (\bibinfo {year} {2023})\BibitemShut {NoStop}%
\bibitem [{\citenamefont {Vaidman}(1994)}]{1994_366}%
  \BibitemOpen
  \bibfield  {author} {\bibinfo {author} {\bibfnamefont {L.}~\bibnamefont
  {Vaidman}},\ }\bibfield  {title} {\bibinfo {title} {Teleportation of quantum
  states},\ }\href {https://doi.org/10.1103/PhysRevA.49.1473} {\bibfield
  {journal} {\bibinfo  {journal} {Physical Review A}\ }\textbf {\bibinfo
  {volume} {49}},\ \bibinfo {pages} {1473} (\bibinfo {year}
  {1994})}\BibitemShut {NoStop}%
\bibitem [{\citenamefont {Braunstein}\ and\ \citenamefont
  {Kimble}(1998)}]{1998_302}%
  \BibitemOpen
  \bibfield  {author} {\bibinfo {author} {\bibfnamefont {S.~L.}\ \bibnamefont
  {Braunstein}}\ and\ \bibinfo {author} {\bibfnamefont {H.~J.}\ \bibnamefont
  {Kimble}},\ }\bibfield  {title} {\bibinfo {title} {Teleportation of
  continuous quantum variables},\ }\href@noop {} {\bibfield  {journal}
  {\bibinfo  {journal} {Physical Review Letter}\ }\textbf {\bibinfo {volume}
  {80}} (\bibinfo {year} {1998})}\BibitemShut {NoStop}%
\bibitem [{\citenamefont {Marian}\ and\ \citenamefont
  {Marian}(2006)}]{marian2006continuous}%
  \BibitemOpen
  \bibfield  {author} {\bibinfo {author} {\bibfnamefont {P.}~\bibnamefont
  {Marian}}\ and\ \bibinfo {author} {\bibfnamefont {T.~A.}\ \bibnamefont
  {Marian}},\ }\bibfield  {title} {\bibinfo {title} {Continuous-variable
  teleportation in the characteristic-function description},\ }\href@noop {}
  {\bibfield  {journal} {\bibinfo  {journal} {Physical Review A}\ }\textbf
  {\bibinfo {volume} {74}},\ \bibinfo {pages} {042306} (\bibinfo {year}
  {2006})}\BibitemShut {NoStop}%
\bibitem [{\citenamefont {Chizhov}\ \emph {et~al.}(2002)\citenamefont
  {Chizhov}, \citenamefont {Kn{\"o}ll},\ and\ \citenamefont
  {Welsch}}]{chizhov2002continuous}%
  \BibitemOpen
  \bibfield  {author} {\bibinfo {author} {\bibfnamefont {A.}~\bibnamefont
  {Chizhov}}, \bibinfo {author} {\bibfnamefont {L.}~\bibnamefont {Kn{\"o}ll}},\
  and\ \bibinfo {author} {\bibfnamefont {D.-G.}\ \bibnamefont {Welsch}},\
  }\bibfield  {title} {\bibinfo {title} {Continuous-variable quantum
  teleportation through lossy channels},\ }\href@noop {} {\bibfield  {journal}
  {\bibinfo  {journal} {Physical Review A}\ }\textbf {\bibinfo {volume} {65}},\
  \bibinfo {pages} {022310} (\bibinfo {year} {2002})}\BibitemShut {NoStop}%
\bibitem [{\citenamefont {Guerrini}\ \emph {et~al.}(2023)\citenamefont
  {Guerrini}, \citenamefont {Win},\ and\ \citenamefont
  {Conti}}]{guerrini2023photon}%
  \BibitemOpen
  \bibfield  {author} {\bibinfo {author} {\bibfnamefont {S.}~\bibnamefont
  {Guerrini}}, \bibinfo {author} {\bibfnamefont {M.~Z.}\ \bibnamefont {Win}},\
  and\ \bibinfo {author} {\bibfnamefont {A.}~\bibnamefont {Conti}},\ }\bibfield
   {title} {\bibinfo {title} {Photon-varied quantum states: Unified
  characterization},\ }\href@noop {} {\bibfield  {journal} {\bibinfo  {journal}
  {Physical Review A}\ }\textbf {\bibinfo {volume} {108}},\ \bibinfo {pages}
  {022425} (\bibinfo {year} {2023})}\BibitemShut {NoStop}%
\bibitem [{\citenamefont {Blasiak}\ \emph {et~al.}(2007)\citenamefont
  {Blasiak}, \citenamefont {Horzela}, \citenamefont {Penson}, \citenamefont
  {Solomon},\ and\ \citenamefont {Duchamp}}]{blasiak2007combinatorics}%
  \BibitemOpen
  \bibfield  {author} {\bibinfo {author} {\bibfnamefont {P.}~\bibnamefont
  {Blasiak}}, \bibinfo {author} {\bibfnamefont {A.}~\bibnamefont {Horzela}},
  \bibinfo {author} {\bibfnamefont {K.~A.}\ \bibnamefont {Penson}}, \bibinfo
  {author} {\bibfnamefont {A.~I.}\ \bibnamefont {Solomon}},\ and\ \bibinfo
  {author} {\bibfnamefont {G.~H.}\ \bibnamefont {Duchamp}},\ }\bibfield
  {title} {\bibinfo {title} {Combinatorics and boson normal ordering: A gentle
  introduction},\ }\href@noop {} {\bibfield  {journal} {\bibinfo  {journal}
  {American Journal of Physics}\ }\textbf {\bibinfo {volume} {75}},\ \bibinfo
  {pages} {639} (\bibinfo {year} {2007})}\BibitemShut {NoStop}%
\bibitem [{\citenamefont {Chatterjee}\ \emph {et~al.}(2012)\citenamefont
  {Chatterjee}, \citenamefont {Dhar},\ and\ \citenamefont
  {Ghosh}}]{chatterjee2012Nonclassical}%
  \BibitemOpen
  \bibfield  {author} {\bibinfo {author} {\bibfnamefont {A.}~\bibnamefont
  {Chatterjee}}, \bibinfo {author} {\bibfnamefont {H.~S.}\ \bibnamefont
  {Dhar}},\ and\ \bibinfo {author} {\bibfnamefont {R.}~\bibnamefont {Ghosh}},\
  }\bibfield  {title} {\bibinfo {title} {Nonclassical properties of states
  engineered by superpositions of quantum operations on classical states},\
  }\href@noop {} {\bibfield  {journal} {\bibinfo  {journal} {Journal of Physics
  B: Atomic, Molecular and Optical Physics}\ }\textbf {\bibinfo {volume}
  {45}},\ \bibinfo {pages} {205501} (\bibinfo {year} {2012})}\BibitemShut
  {NoStop}%
\bibitem [{\citenamefont {He}\ \emph {et~al.}(2021)\citenamefont {He},
  \citenamefont {Malaney},\ and\ \citenamefont {Burnett}}]{2021_354}%
  \BibitemOpen
  \bibfield  {author} {\bibinfo {author} {\bibfnamefont {M.}~\bibnamefont
  {He}}, \bibinfo {author} {\bibfnamefont {R.}~\bibnamefont {Malaney}},\ and\
  \bibinfo {author} {\bibfnamefont {B.~A.}\ \bibnamefont {Burnett}},\
  }\bibfield  {title} {\bibinfo {title} {Noiseless linear amplifiers for
  multimode states},\ }\bibfield  {journal} {\bibinfo  {journal} {Physical
  Review A}\ }\textbf {\bibinfo {volume} {103}},\ \href
  {https://doi.org/10.1103/PhysRevA.103.012414} {10.1103/PhysRevA.103.012414}
  (\bibinfo {year} {2021})\BibitemShut {NoStop}%
\bibitem [{\citenamefont {Kumar}\ \emph {et~al.}(2019)\citenamefont {Kumar},
  \citenamefont {Singh}, \citenamefont {Bose} \emph
  {et~al.}}]{kumar2019coherence}%
  \BibitemOpen
  \bibfield  {author} {\bibinfo {author} {\bibfnamefont {C.}~\bibnamefont
  {Kumar}}, \bibinfo {author} {\bibfnamefont {J.}~\bibnamefont {Singh}},
  \bibinfo {author} {\bibfnamefont {S.}~\bibnamefont {Bose}}, \emph {et~al.},\
  }\bibfield  {title} {\bibinfo {title} {Coherence-assisted non-{G}aussian
  measurement-device-independent quantum key distribution},\ }\href@noop {}
  {\bibfield  {journal} {\bibinfo  {journal} {Physical Review A}\ }\textbf
  {\bibinfo {volume} {100}},\ \bibinfo {pages} {052329} (\bibinfo {year}
  {2019})}\BibitemShut {NoStop}%
\bibitem [{\citenamefont {Ye}\ \emph {et~al.}(2023)\citenamefont {Ye},
  \citenamefont {Guo}, \citenamefont {Zhang}, \citenamefont {Chang},
  \citenamefont {Xia}, \citenamefont {Xiong},\ and\ \citenamefont
  {Hu}}]{2023_473}%
  \BibitemOpen
  \bibfield  {author} {\bibinfo {author} {\bibfnamefont {W.}~\bibnamefont
  {Ye}}, \bibinfo {author} {\bibfnamefont {Y.}~\bibnamefont {Guo}}, \bibinfo
  {author} {\bibfnamefont {H.}~\bibnamefont {Zhang}}, \bibinfo {author}
  {\bibfnamefont {S.}~\bibnamefont {Chang}}, \bibinfo {author} {\bibfnamefont
  {Y.}~\bibnamefont {Xia}}, \bibinfo {author} {\bibfnamefont {S.}~\bibnamefont
  {Xiong}},\ and\ \bibinfo {author} {\bibfnamefont {L.}~\bibnamefont {Hu}},\
  }\bibfield  {title} {\bibinfo {title} {Coherent superposition of photon
  subtraction- and addition-based two-mode squeezed coherent state: quantum
  properties and its applications},\ }\bibfield  {journal} {\bibinfo  {journal}
  {Quantum Information Processing}\ }\textbf {\bibinfo {volume} {22}},\ \href
  {https://doi.org/10.1007/s11128-022-03658-8} {10.1007/s11128-022-03658-8}
  (\bibinfo {year} {2023})\BibitemShut {NoStop}%
\bibitem [{\citenamefont {Selvadoray}\ and\ \citenamefont
  {Kumar}(1997)}]{selvadoray1997phase}%
  \BibitemOpen
  \bibfield  {author} {\bibinfo {author} {\bibfnamefont {M.}~\bibnamefont
  {Selvadoray}}\ and\ \bibinfo {author} {\bibfnamefont {M.~S.}\ \bibnamefont
  {Kumar}},\ }\bibfield  {title} {\bibinfo {title} {Phase properties of
  correlated two-mode squeezed coherent states},\ }\href@noop {} {\bibfield
  {journal} {\bibinfo  {journal} {Optics communications}\ }\textbf {\bibinfo
  {volume} {136}},\ \bibinfo {pages} {125} (\bibinfo {year}
  {1997})}\BibitemShut {NoStop}%
\end{thebibliography}%

\end{document}